\documentclass[useAMS,usenatbib]{mn2e}
\usepackage{amssymb}
\usepackage{amsmath}
\usepackage{graphics,epsfig,graphicx,color}
\usepackage{epic}
\usepackage{rotating}
\usepackage{subfigure}
\usepackage{setspace}
\usepackage{aas_macros}
\usepackage{natbib}
\usepackage{ulem}
\usepackage{longtable}

\usepackage[T1]{fontenc}
\usepackage{times}

\usepackage{color}
\def\changed{}

 \newcommand{\p}{$^{+}$}
 \newcommand{\m}{$^{-}$}
 
\newcommand{\hp}{$\rm{H}^{+}$}
 \newcommand{\htp}{${\rm{H}_2}^+$}
 \newcommand{\hthp}{{\rm{H}_3}^+}

\newcommand{\cm}{\rm{cm}^{-3}}

\title[Pressure-driven fragmentation of multi-phase clouds]
{Pressure-driven fragmentation of multi-phase clouds at high redshift}
\author[H. Dhanoa, J. Mackey \& J. Yates]
{
  H. Dhanoa$^{1}$\thanks{E-mail: hd@star.ucl.ac.uk (HD); jmackey@astro.uni-bonn.de (JM)},
  J. Mackey$^{2}$,
and J. Yates$^{1}$ \\
  $^{1}$ Department of Physics and Astronomy, University College London, Gower Street, London, WC1E 6BT\\
  $^{2}$ Argelander-Institut f\"ur Astronomie, Auf dem H\"ugel 71, Bonn D-53121, Germany
}
\begin{document}


\pagerange{\pageref{firstpage}--\pageref{lastpage}} \pubyear{2013}

\maketitle

\label{firstpage}

\begin{abstract}
The discovery of a hyper metal-poor star with total metallicity of $\le 10^{-5}$ Z$_\odot$, 
has motivated new investigations of how such objects can form from primordial gas polluted by a single supernova. In this paper we present a shock-cloud model which simulates a supernova remnant interacting with a cloud in a metal-free environment at redshift $z=10$.  Pre-supernova conditions are considered, which include a multiphase neutral medium and H\,\textsc{ii} region. A  small dense clump ($n=100$ cm$^{-3}$), located 40 pc from a 40 M$_\odot$ metal-free star, embedded in a $n=10$ cm$^{-3}$ ambient cloud. The evolution of the supernova remnant
and its subsequent interaction with the dense clump is examined.
We include a comprehensive treatment of the non-equilibrium {\changed hydrogen and helium} chemistry and associated radiative cooling that is occurring at all stages of the shock-cloud model,
covering the temperature range $10-10^9$ K. {\changed Deuterium chemistry and its associated cooling are not included
because the UV radiation field produced by the relic H\,\textsc{ii}
region and supernova remnant is expected to suppress deuterium
chemistry and cooling.}
We find a $10^{3}\times$ density enhancement of the clump (maximum density $\approx 78000$ cm$^{-3}$) within this metal-free model. This is consistent with Galactic shock-cloud models considering solar metallicity gas with equilibrium cooling functions. Despite this strong compression, the cloud does not become gravitationally unstable. We find that the small cloud modelled here is destroyed for shock velocities $\gtrsim 50\,$km\,s$^{-1}$, and not significantly affected by shocks with velocity $\lesssim 30\,$km\,s$^{-1}$.
Rather specific conditions are required to make such a cloud collapse, and substantial further compression would be required to reduce the local Jeans mass to subsolar values.
\end{abstract}


\section{Introduction}
The first galaxies are thought to have formed around redshift $z\ge 10$ when the universe was less than 500 Myrs old.
These nascent environments are considered to be the key sites where the transition from Population III to Population II stars took place. 
A possible fossil from this era is SDSS J102915+172927, which is a low-mass ($M<0.8$ $M_{\odot}$) star with a total metallicity of Z$\,<10^{-5}$\,Z$_{\odot}$ \citep{Caffau11}.  As a result of such low metallicity, it is deduced that the star formed from primordial gas which was polluted by a single supernova. This star has challenged the theory that a critical metallicity is needed to form sub-solar-mass Population II star \citep{Klessen12}. A better understanding of star formation and its feedback effects at high redshifts is extremely important in relation to the formation of such objects.

While it is important to study star formation at very low metallicity
\citep{Nagakura09, Chiaki13}, one cannot evaluate the effects of
tiny metal abundances without also studying primordial gas.
The metal-free problem is the limiting case, and is therefore very
useful as a baseline study for comparison to later calculations for gas
that is polluted by trace amounts of metals.
It is also very interesting in its own right, because we still do not
know if stars of mass $<1$ M$_\odot$ can form at zero metallicity (see e.g.\
the interesting proposal presented by \citet{Stacy14}). 

Here we examine the shock-cloud interaction model 
developed by \citet{MacBroHer03},
in which shock compression and subsequent cooling can
decrease the Jeans mass in primordial gas, thereby forming lower-mass
stars than would form without the shock collision.
Our work is the first to investigate this problem
with
detailed multi-dimensional simulations for metal-free gas.
Radiative cooling is the critical factor in promoting hydrodynamic and gravitational instabilities. Therefore in this paper we focus on the non-equilibrium cooling that dominates this system. This can only be captured correctly by including non-equilibrium chemistry (linked to thermal models) for the full temperature range associated with a supernova shock model. {\changed We have focused on hydrogen and helium chemistry because we expect
that the environment surrounding a progenitor Population III star is
dominated by H$_2$ cooling.
Both the relic H\,\textsc{ii} region and the supernova remnant are
sources of diffuse UV radiation that suppresses HD cooling, so we have
not included deuterium chemistry in this work. \citet{WolHai11} showed that HD cooling is strongly
suppressed by UV radiation fields that are up to five orders of
magnitude weaker than what is required to suppress H$_2$ cooling. }

\citet{Kitayama05} and \citet{Vasiliev08} highlighted the important link between the radial distribution of primordial gas prior to the supernova explosion and the subsequent evolution of the supernova remnant, and therefore the formation of extremely metal-poor stars. Consequently, we include both the H\,\textsc{ii} region and neutral medium, to obtain a realistic supernova shell evolution. Once the supernova shock begins to travel within neutral matter, it interacts with a multi-phase medium \citep{Reach05}, which cannot be characterised by a single density.  \citet{Greif08} have found that turbulence driven by cold accretion onto a protogalaxy produces a primordial interstellar medium with a large range of densities and temperatures. The pressure-driven compression and fragmentation of dense clumps found in this neutral matter could be a possible site for low-mass star formation.

At present most supernova shock models for the early universe only include non-equilibrium cooling for temperatures below $10^4$ K and focus on the fragmentation of the supernova shell itself. \citet{Machida05} were the first to investigate primordial low-mass star formation at high redshift via this method. The authors included non-equilibrium cooling from H$_2$ and HD molecules, coupled to a semi-analytic dynamic model. They found that shell fragmentation was possible for explosion energies $\ge 10^{51}$ erg and ambient density $n>3$ cm$^{-3}$. The contraction of the fragments was studied, and the Jeans mass was reduced to $\sim$1 M$_\odot$. \citet{Nagakura09} extended this model to include metal-line cooling for low-metallicity gas coupled to a 1D hydrodynamic code. They use linear perturbation analysis of the expanding shell to constrain the criteria for fragmentation and found that there is little dependency on metallicity in the range 10$^{-4}-10^{-2}\, Z_{\odot}$. Compared to \citet{Machida05}, they found that fragmentation only occurred in higher ambient uniform densities ($n\ge 100$ cm$^{-3}$ for a $10^{51}$ erg explosion and $n\ge 10$ cm$^{-3}$ for a $10^{52}$ erg explosion), and eventually form fragments of mass $10^2-10^3$ M$_\odot$.

\citet{Chiaki13} developed a 1D supernova model that considers a gas with metallicity $10^{-5}\,\rm{Z}_\odot$. The authors include metal-free non-equilibrium chemistry for temperatures below $10^4$ K, with separate calculated rates for metal-line cooling. However, above $10^4$ K the authors utilise the collisional ionization equilibrium cooling function by \citet{Sutherland93}. The authors find that the supernova shell becomes gravitationally unstable for a wide range of explosion energies ($10^{51}-3\times 10^{52}$ erg) and ambient uniform densities ($n\ge 10$ cm$^{-3}$). The thermal evolution of a shell fragment was followed
using a one-zone model (a point calculation) which includes low-metallicity chemistry and dust cooling. They expect the fragment to evolve into a high density core ($10^{13}$ cm$^{-3}$), which will eventually form multiple clumps of mass $0.01 -0.1$ M$_\odot$.

Using a one-zone model, \citet{MacBroHer03} modelled an equilibrium primordial gas cloud that is shocked by a supernova. The shocked cloud is heated to a higher entropy state and it is assumed to cool isobarically back to its original equilibrium temperature, but now at a much higher density than before. In this way the Jeans mass of the gas could be reduced by a large factor, allowing much lower-mass stars to form. This argument also applies to smooth ISM distributions, as discussed above \citep{Machida05, Nagakura09, Chiaki13}, as long as isobaric conditions hold in the decelerating shell. 

The one-zone model of \citet{MacBroHer03} also crucially depends on the isobaric assumption to increase the gas density in the cooling cloud. In reality, however, pressure is a decreasing function of time in a supernova remnant, because the explosion is (by definition) vastly over-pressurised compared to its surroundings. As long as the expansion timescale of the supernova $t_\mathrm{exp}=R_\mathrm{sh}/\dot{R_\mathrm{sh}}$ (where $R_\mathrm{sh}$ is the shock radius and $\dot{R_\mathrm{sh}}$ its velocity) is short compared to the local timescale for gravitational effects (i.e~the free-fall time $t_\mathrm{ff}=1/\sqrt{G\rho}$, where $\rho$ is the gas density and $G$ the gravitational constant) then the time-dependence of the external pressure is an important part of the solution.
The passage of a strong shock through a dense cloud can also have catastrophic consequences for the cloud \citep{KleMcKCol94} through turbulent hydrodynamic instabilities.
Both of these considerations are best addressed with multi-dimensional hydrodynamic simulations and cannot be captured in one-zone models.

\citet{Melioli06} investigated star formation triggered in the Galactic environment, via the interaction of a supernova shell and molecular cloud. The authors produce constraints on cloud collapse (and therefore possible star formation) in the `supernova remnant radius vs. cloud density' parameter space. This was achieved by an analytic study comparing the gravitational free-fall time and destruction time scale of the cloud (which depends on a number of parameters including radiative cooling). By running a suite of 3D hydrodynamic simulations, they were able to confirm that these numerical models were consistent with their analytic constraints. The authors recognise that using an approximate polytropic pressure equation to represent radiative cooling maybe an over simplification and more realistic cooling functions are required.


\citet{Johansson13} have concentrated on the compression of smaller clouds (radius $\sim$ 1 pc) found in the local interstellar medium as a method of triggered star formation. Their MHD simulations (without self-gravity) concentrate on the radiative interaction between the shock and the cloud. The cooling function utilised is a piecewise power-law given by \citet{Sanchez02} and \citet{Slyz05}, and assumes collisional ionization equilibrium. They find that the cloud fragments into small dense cool clumps and do not become Jeans unstable. Importantly they find that initial density enhancements within the cloud can increase by a factor of $10^3 -10^5$, which eventually relaxes to a final density enhancement of $10^2-10^3$. This is consistent with results by \citet{Vaidya13}, who have a similar model which includes self gravity. They find that gravity does not contribute to the large increase in density but plays an important role by preventing the re-expansion of the high density region.

These studies have highlighted that radiative cooling is a crucial process in the interaction between shocks and clouds. In this paper we simulate a supernova exploding in a metal-free environment and include the non-equilibrium radiative cooling that occurs at all stages of its evolution and subsequent collision with a multiphase neutral cloud. The diffusion of the metals is neglected and the system is approximated by primordial chemistry. Hence we present a model which includes the non-equilibrium metal-free chemistry and its associated cooling for the evolution of a supernova shell and its subsequent interaction of a small dense clump embedded in a neutral cloud at redshift $z=10$. In section \S\ref{sec:model} we outline how the initial conditions are generated by the pre-supernova model, and introduce the chemo-dynamic modelling of the supernova remnant. The results describing the generation of the pre-supernova model, the 1D Supernova model and the 2D interaction of the clump and shock, are presented in section \S\ref{sec:results}. Finally, in sections \S\ref{sec:dicusssion} and \S\ref{sec:conc} we discuss our findings and give a summary of the conclusions.

\section{Methods and initial conditions}\label{sec:model}
We have modelled the interaction of a supernova shell with a dense clump in three stages:
\begin{enumerate}
\item the pre-supernova phase, where the dynamical effects of photoionization heating from the star are modelled;
\item the post-supernova phase, where the supernova blast wave expands into the relic H\,\textsc{ii} region left by the star; and
\item the shock-cloud interaction, where the expanding supernova shell compresses a dense cloud.
\end{enumerate}
The first two stages are simulated in one dimension with spherical symmetry, whereas the third stage is simulated in two dimensions with rotational symmetry using a. This is because compression and fragmentation of the clump cannot be captured within 1D models. However, it is possible to achieve a good representation of the evolution of the supernova remnant in 1D models, assuming that the shell has not interacted with any dense clumps \citep{Jun96}.

For the 1D simulations we use reflective boundary conditions at the
origin (imposed by the symmetry of the problem), and a zero gradient
outflow condition at the large radius boundary.
For the 2D simulations with cylindrical coordinates $(R,z)$ we use a
reflective boundary at $R=0$ (again imposed by symmetry) and zero
gradient at $R=R_\mathrm{max}$, an inflow boundary at
$z=z_\mathrm{min}$, and zero gradient at $z=z_\mathrm{max}$.
The inflow boundary condition is justified because the post-shock flow
variables change slowly for $\approx 5-7$ pc behind the blast wave (see
Fig.~2).

{\changed As argued in the Introduction, we do not expect HD cooling to be
important because the supernova shell and dense clump are exposed to
UV radiation from the nearby relic H\,\textsc{ii} region and expanding
supernova remnant.
HD cooling is much more readily suppressed by UV radiation than H$_2$
cooling \citep{WolHai11}, so we focus here only on the
hydrogen and helium chemistry and cooling }
\subsection{Pre-supernova phase}

We use the radiation-magnetohydrodynamics code \textsc{pion} \citep{Mackey10, MacLim11} for the simulations presented here, first in 1D with spherical symmetry and later in 2D with rotational (axi-)symmetry. \textsc{pion} uses an explicit, finite-volume, integration scheme that is accurate to second order in time and space \citep{Fal91}. Here only the Euler equations of hydrodynamics are solved, together with the ionization rate equation of hydrogen and associated non-equilibrium heating and cooling processes.
The microphysical processes of ionization, recombination, heating and cooling are coupled to hydrodynamics using Algorithm 3 in \citet{Mac12}.

We consider a metal-free star exploding in a small galaxy at redshift $z=10$, sweeping up the ambient medium to form an expanding shell.
The simplified initial condition consists of a uniform neutral interstellar medium with hydrogen number density $n=10\,\mathrm{cm}^{-3}$.
Into this we place a dense cloud with (uniform) number density $n=100\,\mathrm{cm}^{-3}$, radius $r_\mathrm{c}=1.3$\,pc, and located at $r=40$\,pc from the star (which is at the origin).
The gas is comprised of atomic hydrogen and helium (number density ratio of 1.00:0.08) and is cooled via atomic processes.
We assume the star has formed in a sufficiently large galaxy that gravitational potential gradients can be neglected in the hydrodynamical evolution of the system.
This is the simplest possible model for feedback from the massive star to a nearby cloud.

For the star's properties we take the 40 M$_\odot$ metal-free model from \citet{Schaerer02} with no mass loss.
This has a lifetime of 3.86\,Myr, an effective temperature $T_\mathrm{eff}=10^{4.9}$\,K, and a time-averaged H-ionising photon luminosity $Q_0=2.47\times10^{49}$\,s$^{-1}$.
For simplicity we distribute these photons according to a blackbody spectrum with the star's $T_\mathrm{eff}$.
We ignore any post main sequence evolutionary effects because this comprises a small fraction of the star's life, and because the evolution is very uncertain.
This model in \citet{Schaerer02} also remains relatively blue for its full lifetime, thus supporting our approximation of excluding a red supergiant phase.


\subsection{Supernova Remnant phase}
A supernova remnant is dominated by non-equilibrium cooling, therefore we developed a microphysics module which links the non-equilibrium chemistry and its associated cooling. This was accomplished by solving the following set of equations:
\begin{eqnarray}
\frac{\partial{E}}{\partial{t}}&=&-\Lambda(\Sigma x_m,\rho,\mbox{T}) + \Gamma(\Sigma x_n,\rho,\mbox{T})\\
\frac{\partial x_i}{\partial t}&=& C_i \left( x_j,\rho, \rm{T}\right) -D_i\left( x_j,\rho,\rm{T}\right) x_i\label{chemical_ODE_eqn}
\end{eqnarray}
where $E$ is the internal energy density (in erg\,cm$^{-3}$),
$\Lambda$ is the cooling function of the gas (in erg\,cm$^{-3}$\,s$^{-1}$),
$\Gamma$ is the heating function of the gas  (in erg\,cm$^{-3}$\,s$^{-1}$), $x_i$ is the fractional abundance of a chemical species, $i$, for a total number of chemical species $N_\mathrm{s}$,
T is the temperature of the gas (K), $\rho$ is the total mass density of the gas (g cm$^{-3}$), C is the  formation rate of the species and D is the destruction rate of the species.
We use a chemical network of 11 species (H, He, H$_2$, H$^+$, H$_2^+$, H$_3^+$, HeH$^+$, He$^+$, He$^{++}$, H$^-$ and e$^-$) and 42 reactions.
The chemical rates cover the temperature range $10-10^{9}$ K, which are described in appendix \ref{appen:chem}.
The atomic species and electron fraction are treated numerically as conservation equations.

  The supernova is modelled by injecting thermal energy, not kinetic (i.e.~we ignore the free-expansion phase). Therefore at very early times the newly shocked gas has an artificially high temperature (T$>10^9$ K),  and at these temperatures we utilise the value of the reaction rates at $10^9$ K. To avoid artificial overcooling at early times, we only switch on the cooling when the gas adiabatically cooled down to $10^8$ K. The thermal model includes atomic cooling \citep{Fukugita94,Hummer94}, Bremmstrahlung cooling  \citep{Hummer94, Shapiro87}, inverse Compton scattering \citep{Peebles71} and molecular line cooling from H$_2$, H$_2^+$ and H$_3^+$ \citep{Glover08, Hollenbach79, Glover09}. The heating processes included in the model are CMB heating (assumed equal to $\Lambda(\rm{T}_{CMB})$) and cosmic ray heating \citep{Glover07}. We set the cosmic ray ionization rate at $\zeta=10^{-18}\,\rm{s}^{-1}$  assuming the supernova remnant to be their source. The chemical model, together with tests of the chemistry and dynamics, are presented in the appendices.

Both the 1D pre-supernova and post-supernova models consist of 5120 grid points to cover a 50 pc range, and are run until it the SN shell reaches 4 pc from the clump centre.
The output of this phase ii model (both chemical and dynamic properties)
is then mapped onto a 2D grid which covers an area of
$9.60\times3.20$ pc ($480\times160$ grid zones, 0.02 pc per zone)
to study the shock-cloud interaction (phase iii).

\section{Results} \label{sec:results}
\begin{table}
\begin{center}
\begin{tabular}{l c}
\hline
Parameters&\\
\hline\hline
Shell thickness & 0.08 pc\\
Maximum shell density&1976 cm$^{-3}$ \\
Minimum shell temperature& 920 K\\
Shell velocity & 39 km s$^{-1}$\\
Clump radius & 1.3 pc\\
Maximum clump density & 104 cm$^{-3}$\\
Minimum clump temperature& 872 K \\
\hline
\end{tabular}
\caption{Initial conditions of 2D model}\label{IC_table}
\end{center}
\end{table}

\subsection{Pre-supernova phase}

\begin{figure}
 \includegraphics[width=0.4\textwidth]{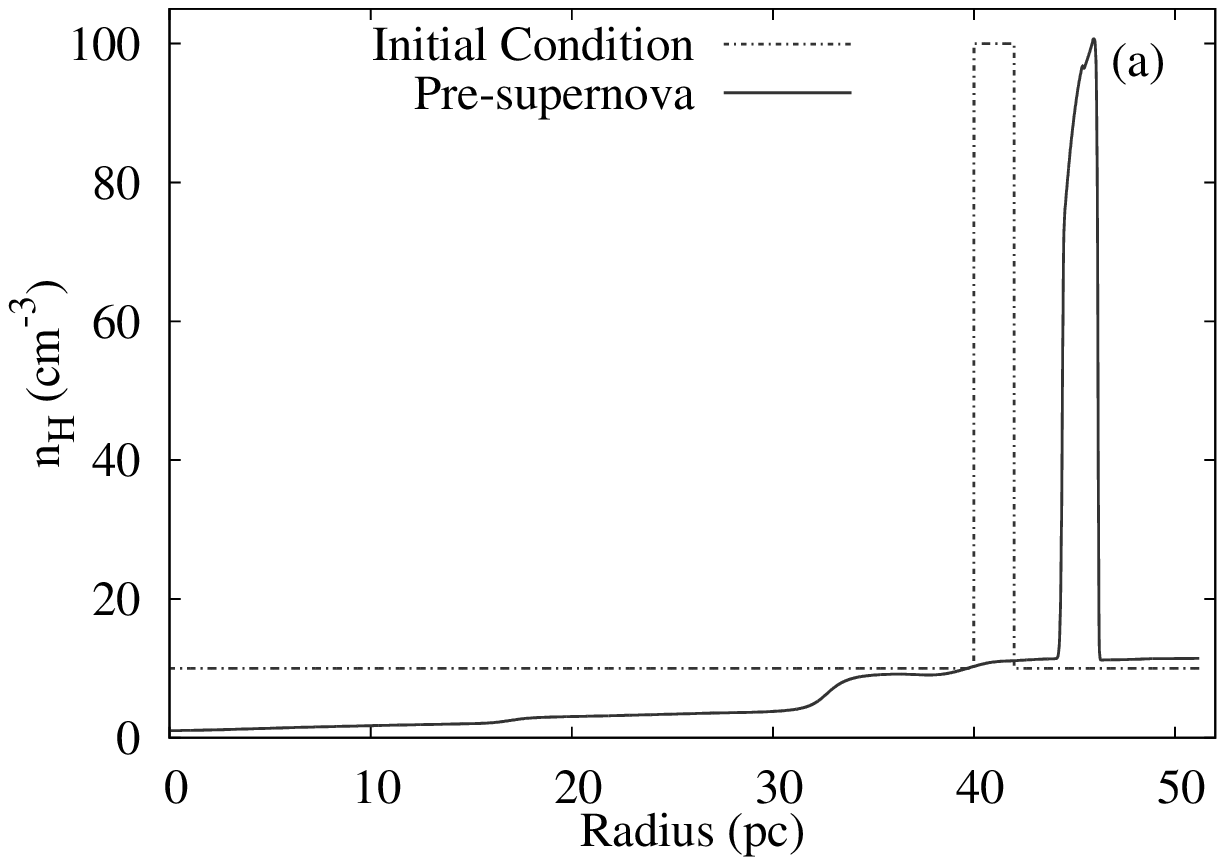}
 \includegraphics[width=0.4\textwidth]{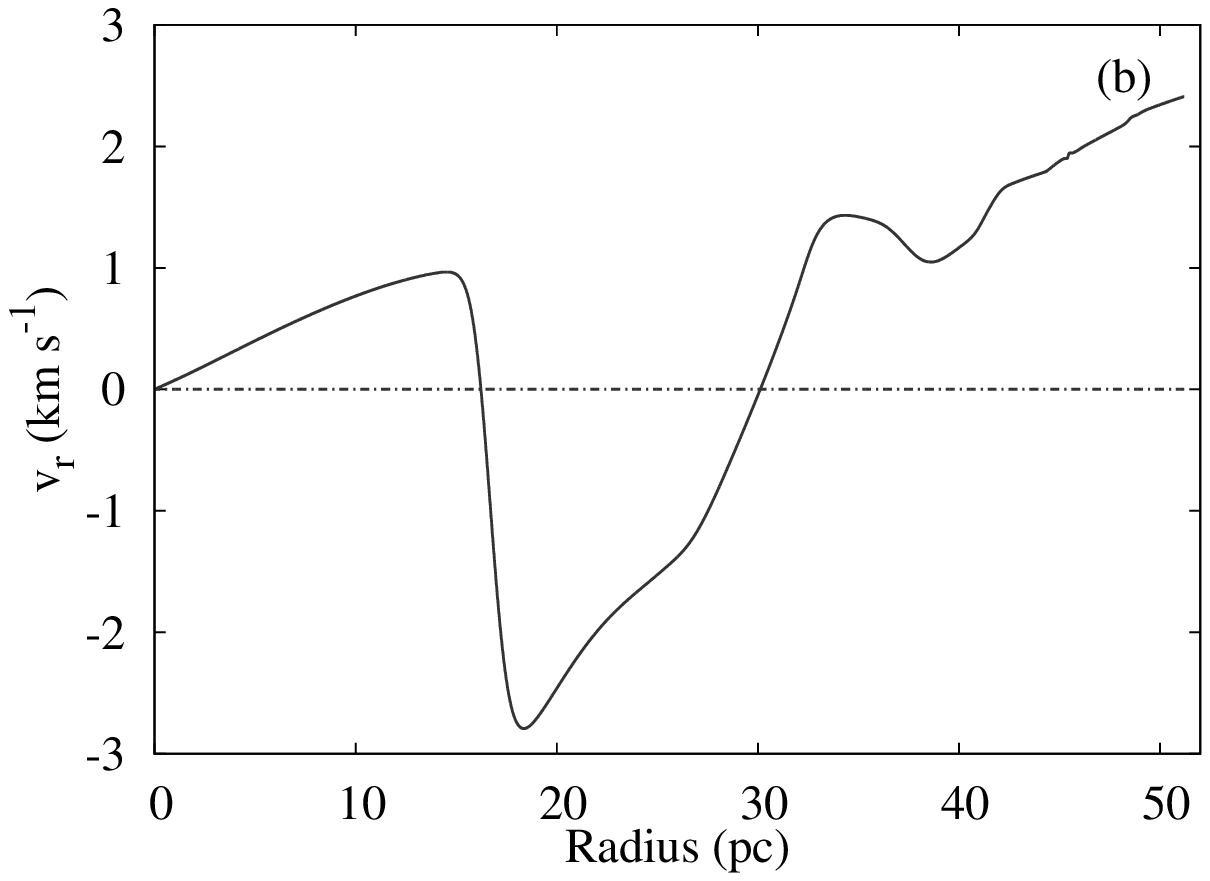}
 \includegraphics[width=0.4\textwidth]{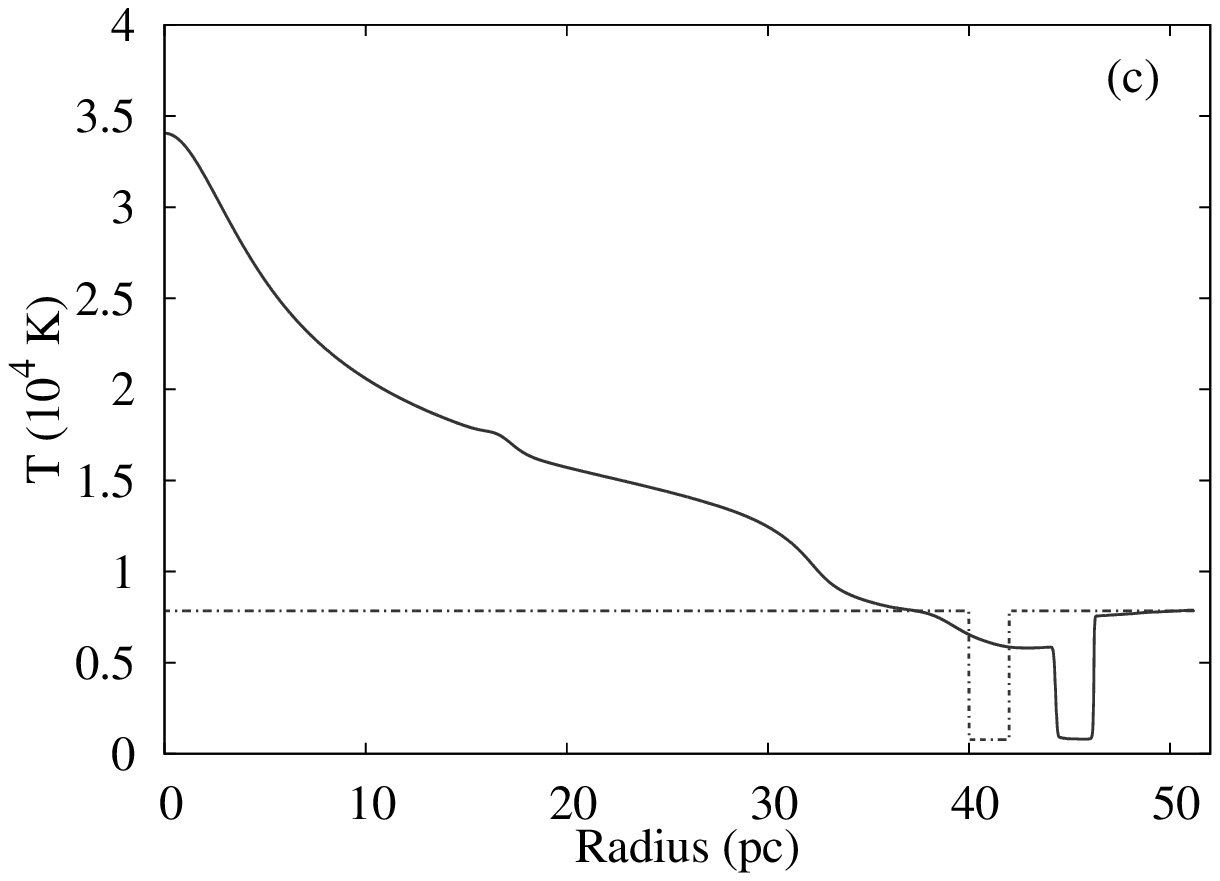}
 \includegraphics[width=0.4\textwidth]{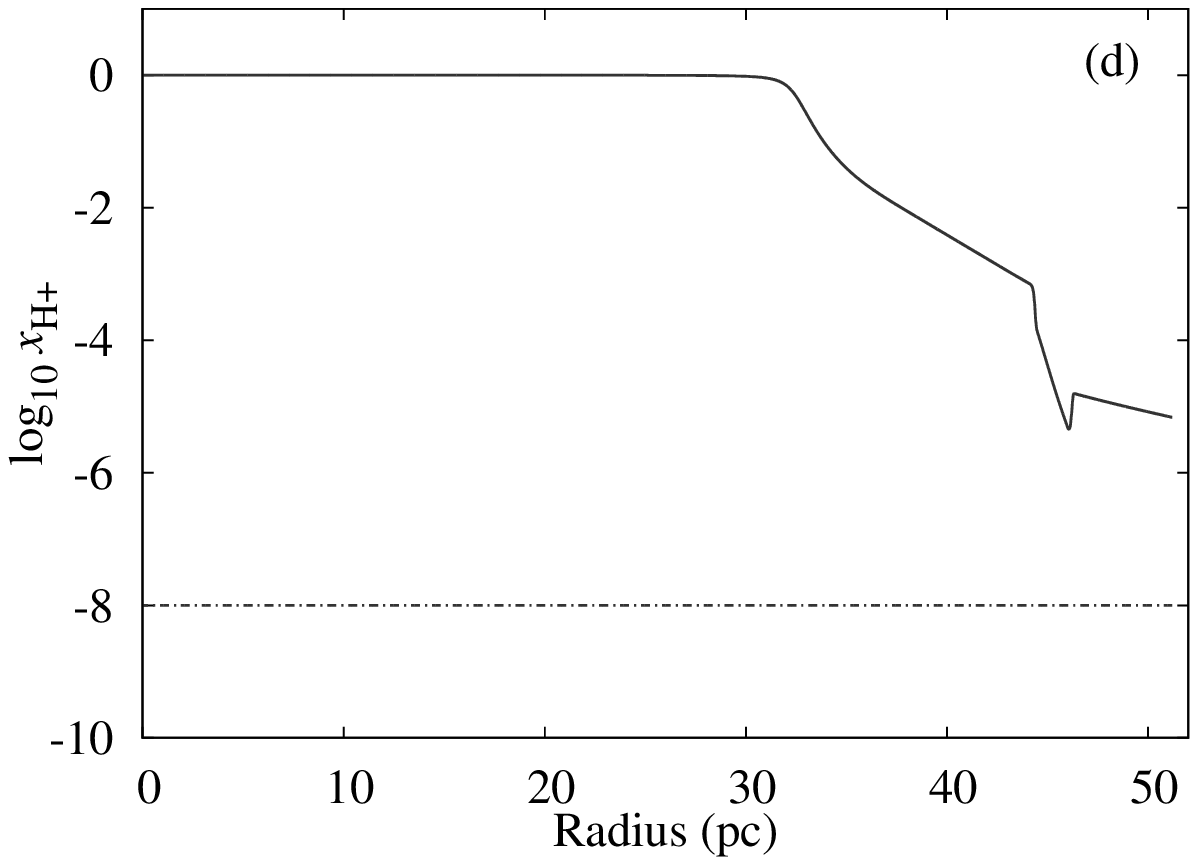}
 \caption{
   Plots of gas number density (a), velocity (b), temperature (c), and H$^{+}$ fraction (d) as a function of distance from the star.
   The dashed lines show the initial conditions and the solid lines the conditions at the pre-supernova stage.
 }
 \label{fig_10cc_HII}
\end{figure}

The radial profile of the initial conditions and the pre-supernova ISM are plotted in Fig.~(\ref{fig_10cc_HII}).
The gas density inside the photoionised H\,\textsc{ii} region ($r<33$\,pc) has decreased compared to the initial conditions (to close to $n=1\,\mathrm{cm^{-3}}$) because photoheating has driven its expansion. In this phase we only include atomic cooling, we assume that the H$_2$ within the gas has been destroyed as a result of Lyman-Werner radiation from the star. The shocked neutral ISM has only weak atomic coolants and so has not formed a shell, and remains very close to the initial ISM density.
The cloud (or in 1D a shell) has been pushed outwards by the H\,\textsc{ii} region expansion, and is moving out at $v\approx2$\,km\,s$^{-1}$ (Fig.~\ref{fig_10cc_HII}b).
The wave reflected back inwards is driving the negative velocity seen between $16<r<30$\,pc, and this is a transient feature imposed by the assumed spherical symmetry (which forces waves to reflect back and forth between the origin and any strong discontinuities).
It has little effect on the overall solution except to marginally increase the density in this radius range.
The temperature profile of the H\,\textsc{ii} region is typical of that produced by hot stars in metal-free gas \citep{IliCiaAlvEA06}.

\subsection{Supernova Remnant phase}
\begin{figure}
 \includegraphics[width=0.4\textwidth]{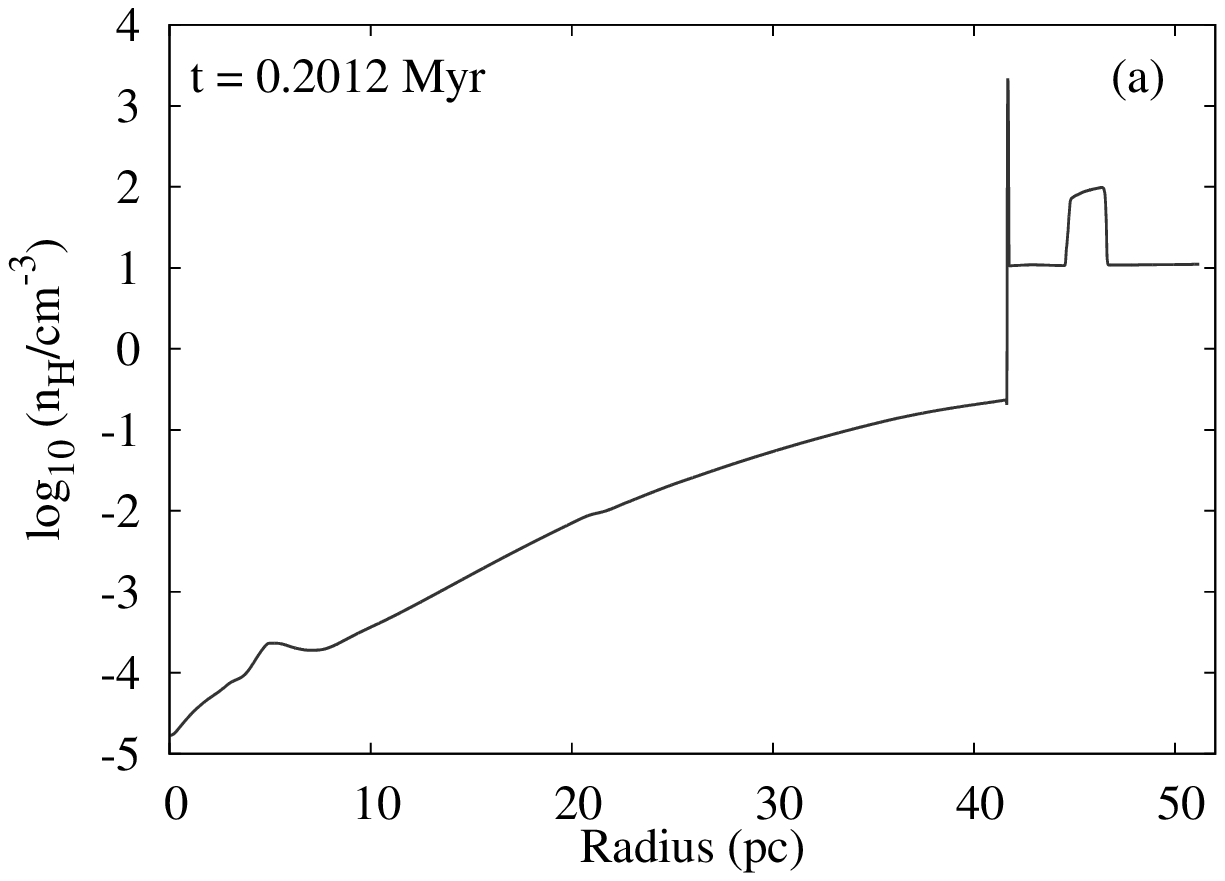}
 \includegraphics[width=0.4\textwidth]{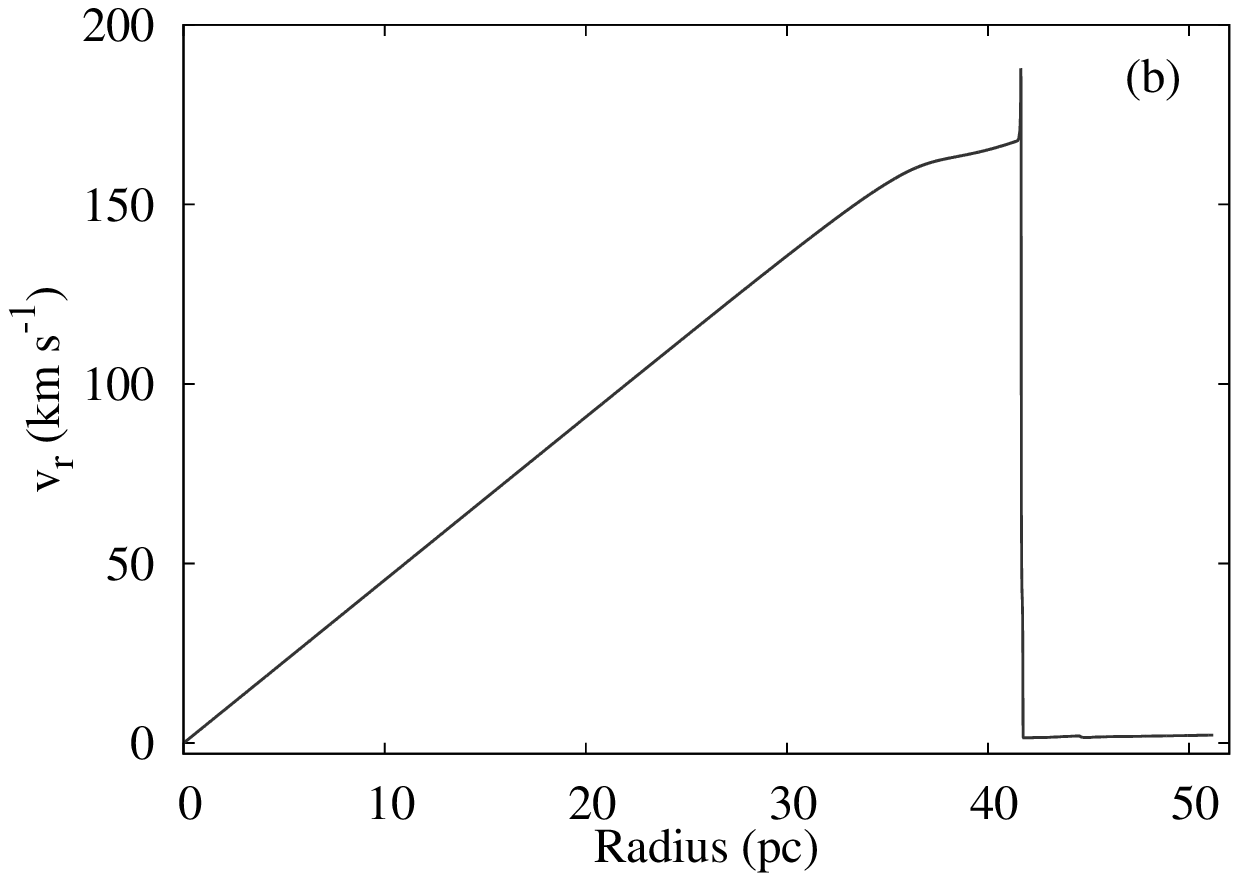}
 \includegraphics[width=0.4\textwidth]{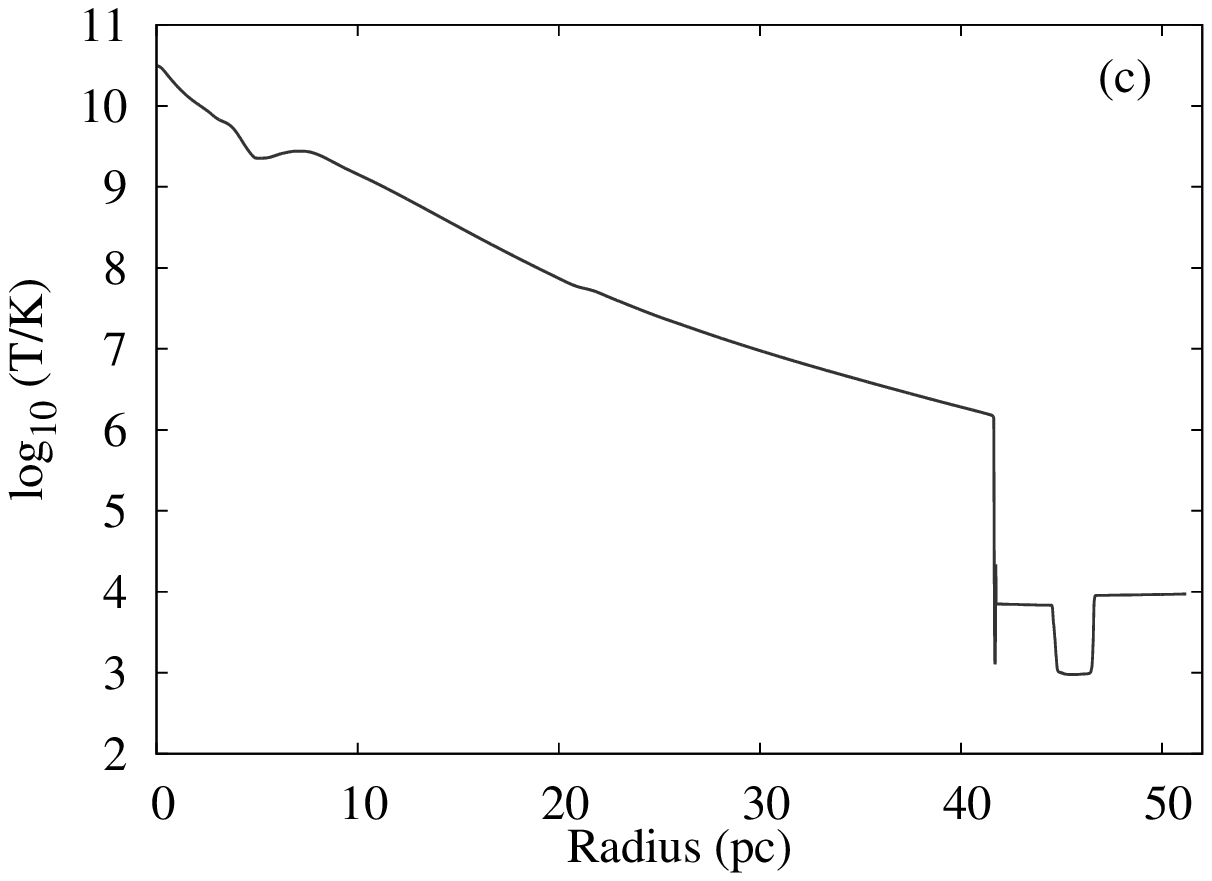}
 \includegraphics[width=0.4\textwidth]{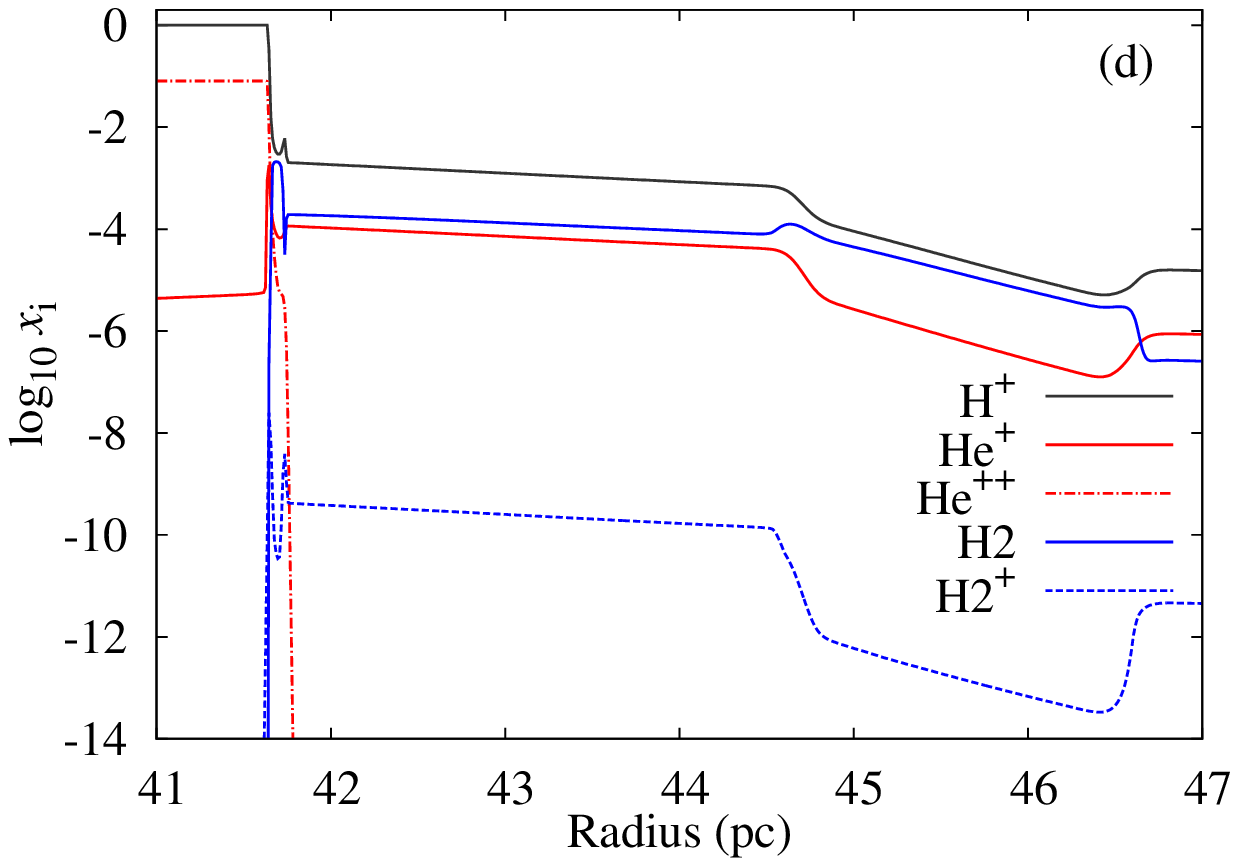}
 \caption{
Gas number density (a), expansion velocity (b), temperature (c), 
and species fractions (d) as a function of distance from the star for the 1D
post-supernova evolution, at $t=0.2012$\,Myr after the supernova explosion.
    Note that panel (d) has a different $x$-axis to the other panels, 
zoomed in to show only the chemistry of the supernova shell and the
overdense cloud (smaller and larger radii show little variation).
    The supernova shell is at $r\approx41.7$\,pc, and the overdense 
cloud at $r\approx44.6-46.6$\,pc.
}
 \label{fig_10cc_1DSN}
\end{figure}

\begin{figure}
  \centering
  \subfigure{
    \label{fig_JM_DensTemp1}
    \includegraphics[width=0.49\textwidth]{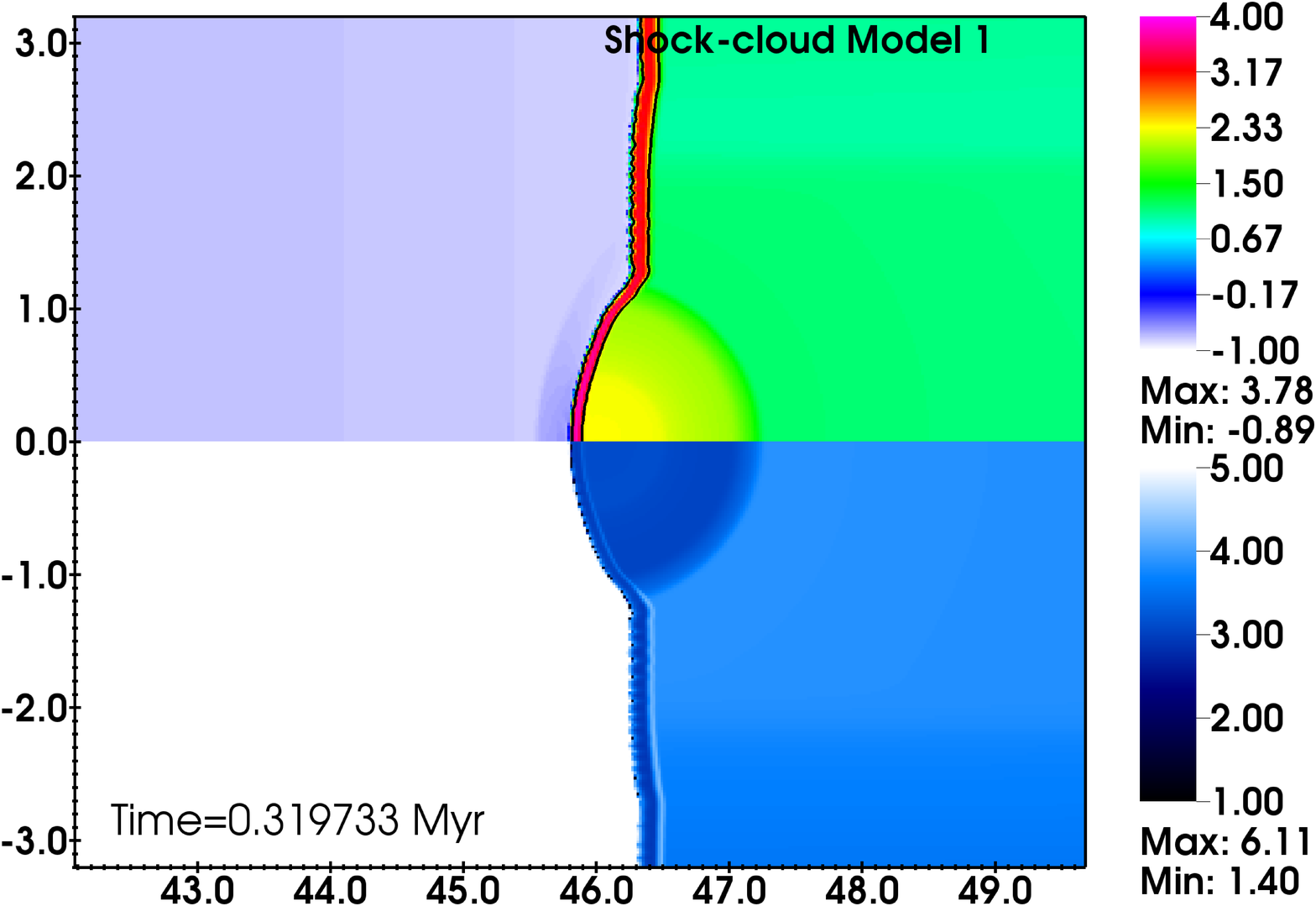}
  }
  \subfigure{
    \label{fig_JM_DensTemp2}
    \includegraphics[width=0.49\textwidth]{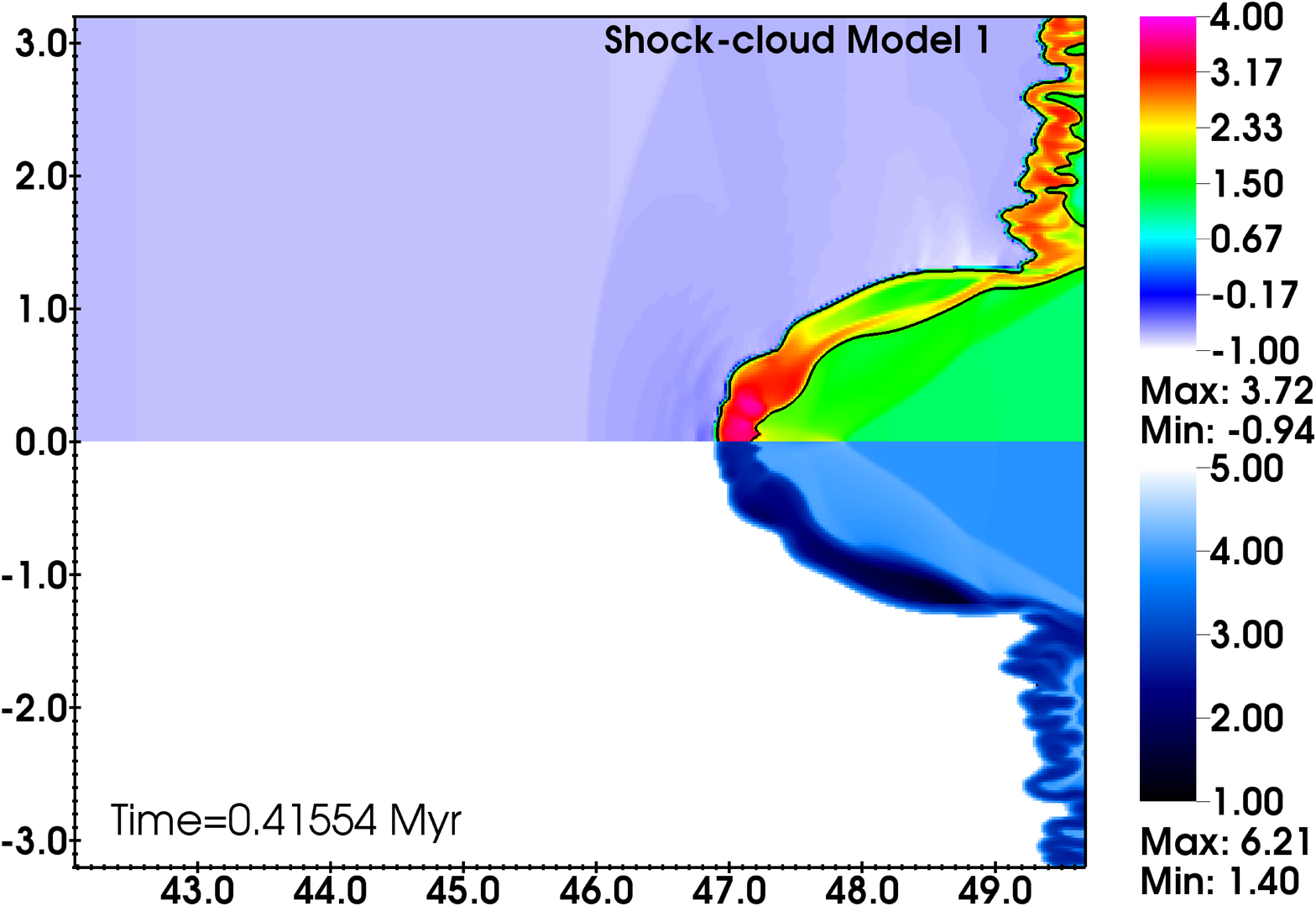}
  }
  \subfigure{
    \label{fig_JM_DensTemp3}
    \includegraphics[width=0.49\textwidth]{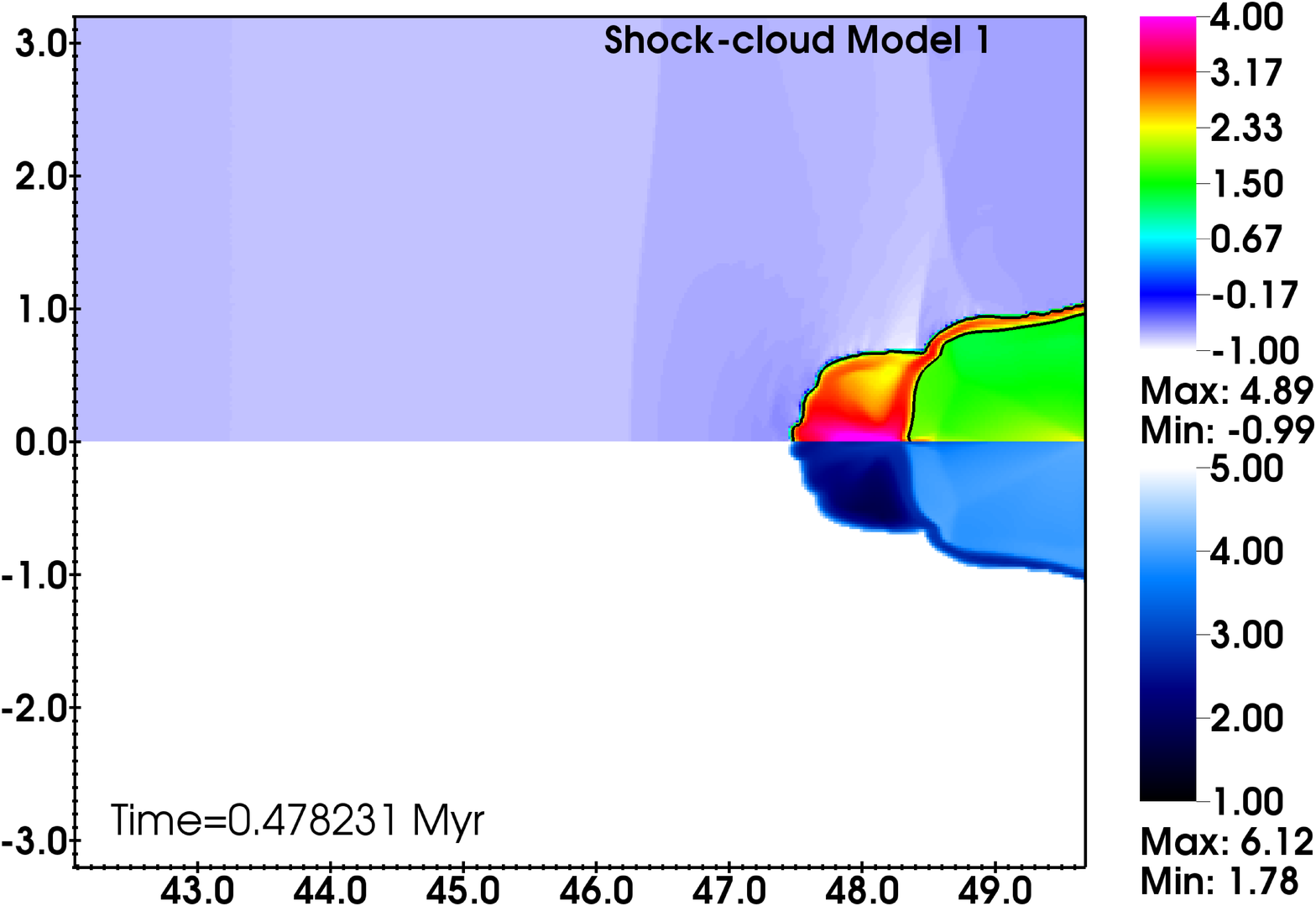}
  }
  \caption{
    Log of H number density ($\log_{10}\,(n_\textsc{h}/$cm$^{-3}$), colour scale) is plotted on the upper half-plane, and Log of temperature on the lower half-plane (blue scale, in Kelvin), with a single black contour line overplotted on the upper half-plane showing where the H$_2$ fraction equals 0.001.
   The panels show an early time as the cloud is being shocked (top), after the shock has passed through the cloud (centre), and after the cloud has been compressed and accelerated by the shock (bottom).
    The x-axis shows distance from the star in parsecs, and the y-axis shows radial distance from the axis of symmetry of the 2D calculations (the lower half-plane is a reflection of the simulation domain to negative values).
  }
  \label{fig_10cc_SN}
\end{figure}

\begin{figure}
 \includegraphics[angle=0,width=0.49\textwidth]{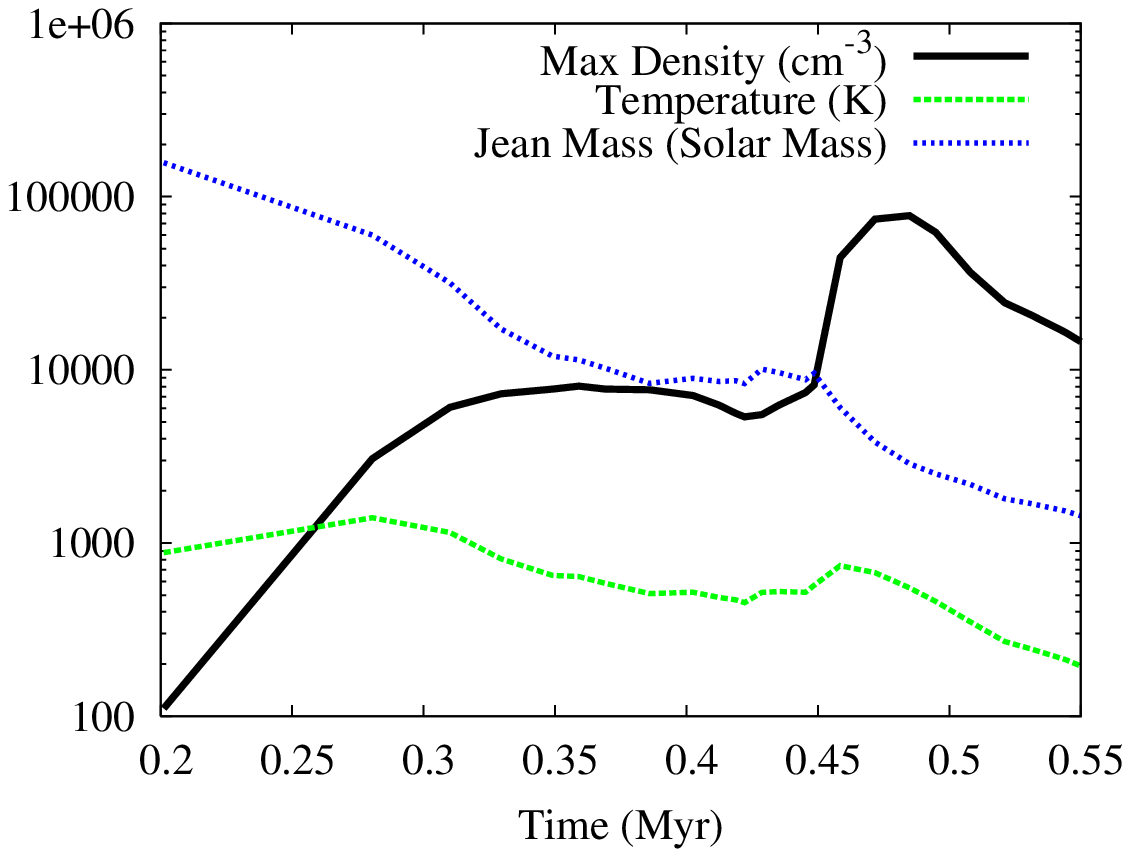}
 \includegraphics[angle=0,width=0.49\textwidth]{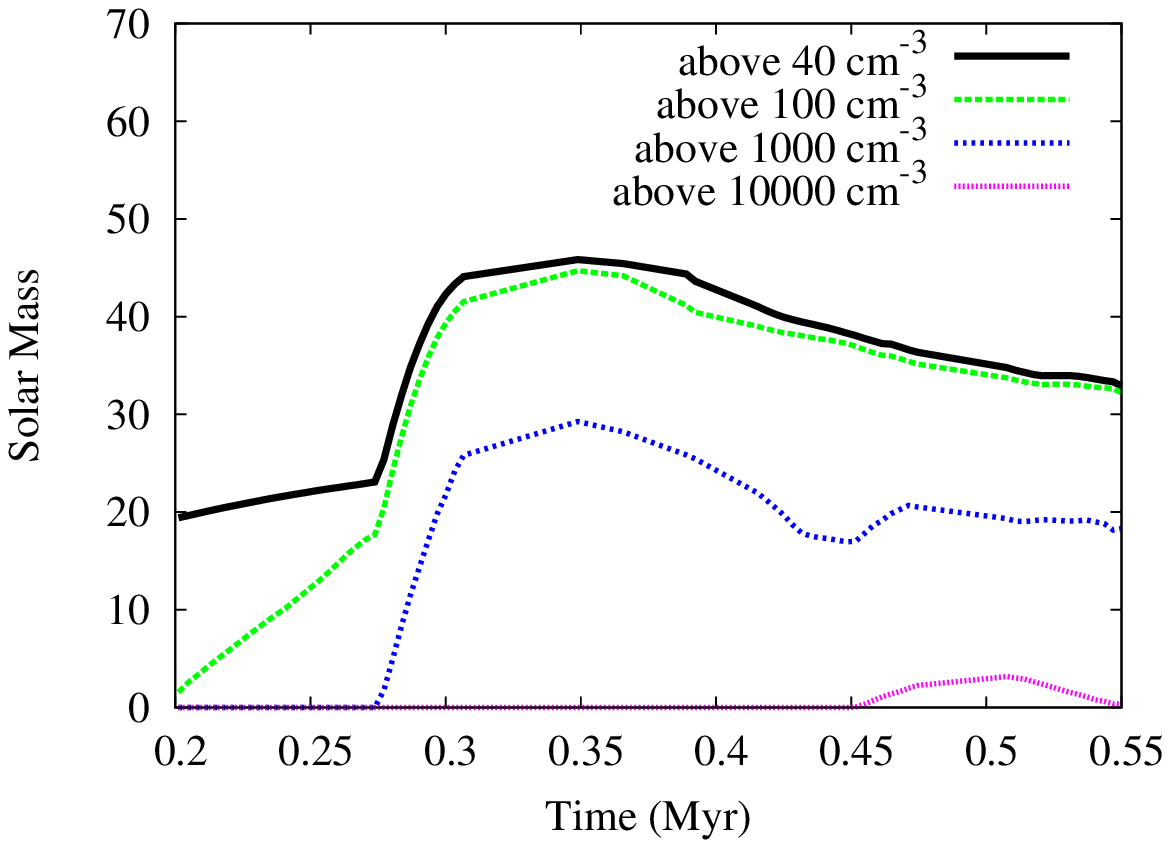}
  \caption{ The upper plot displays the maximum density of within the clump as the shock passes through, along with the temperature of the maximum density point and associated Jeans mass. The lower plot displays the mass within the clump as a function of different densities.
   }
  \label{fig_clump}
\end{figure}

The output from the pre-supernova model is utilised as the initial conditions of the 1D supernova model. The clump has been moved to 45 pc due to  the weak shock driven by dynamical expansion of the H\,\textsc{ii} region (Figure~\ref{fig_10cc_HII}a). When mapping the chemical species, we assume the percentage of ionised hydrogen and helium (He$^+$) are equal, and the initial molecular fractions are set to zero. A $10^{52}$ erg explosion is initiated and a shell starts to form at around $27$ pc.  After 0.2012\,Myr the supernova shock is well into the radiative phase, so a thin shell has formed that is about 200$\times$ denser than the pre-shock gas.
This agrees well with the isothermal shock jump conditions, where the overdensity is equal to the Mach number ($\mathcal{M}$) squared.
In the shell, the isothermal sound speed is $a\approx2.5$\,km\,s$^{-1}$, so $\mathcal{M}^2 \approx (39/2.5)^2\approx240$.
This is also similar to the maximum overdensity obtained from the test calculation in Appendix~\ref{app:SNtest}. In the interior of the supernova remnant the usual Sedov-Taylor solution remains imprinted on the fluid quantities: the density and velocity tend to zero at the origin, and the temperature increases to maintain the constant interior pressure.
The molecular fractions are all negligible in the hot interior, and have a maximum in the shocked shell because here the density is highest but there is also still a non-negligible electron fraction from heating in the shell's forward shock.
The maximum H$_2$ fraction in the shell is $x(\mathrm{H}_2)\approx0.002$, in agreement with previous work \citep{Machida05}.

The 1D supernova model is terminated when the shell reaches 41.9 pc (before it collides with the clump) and the output of this simulation (Figure  \ref{fig_10cc_1DSN}) is mapped onto a 2D axisymmetric grid. The initial conditions for the 2D model are outlined in Table \ref{IC_table}. The supernova shell is already travelling within the neutral ambient medium
and is proceeding towards a dense spherical clump ($\sim19$ M$_\odot$) at a velocity of 39 km s$^{-1}$.
The clump centre is 46 pc from the progenitor star.
Figure \ref{fig_10cc_SN} displays the evolution of the clump as the supernova shell collides and compresses it.
The upper half plane of the plots display the log of the number density (log$_{10}$ $n_{\rm{H}}$/cm$^{-3}$) and corresponding lower half plane plots log of gas temperature reflected about the axis of symmetry. The black contour shows where the H$_2$ fraction equals $10^{-3}$.

After 0.31 Myr the shock has passed through half of the clump (upper plot in Figure \ref{fig_10cc_SN}), we can see from Figure \ref{fig_clump} the maximum density of clump is  $\sim6000$ cm$^{-3}$ with an associated temperature of $\sim1000$ K.
The supernova shell has passed through the clump completely by 0.41 Myrs (middle plot in Figure \ref{fig_10cc_SN}), and due to the decline in pressure the maximum density has decreased to $\sim 5200$ cm$^{-3}$.
The shock has caused an increase in free electrons, which catalyse the formation of H$_2$. Hence the temperature of the high density gas has decreased to $\sim 400$ K. As the supernova shell passes through and around the clump, the region
of strong shear at the clump's edge undergoes adiabatic expansion and
cools to close to the CMB temperature (middle panel of Fig.~3).  This is
not radiative cooling; the minimum temperature of the densest gas (with
the strongest cooling) is $\sim400$ K. The clump reaches its maximum density of $\approx78000$ cm$^{-3}$ around
0.47 Myrs after the initial supernova explosion (bottom panel of Fig.~3).
Again we see the densest gas is not the coldest gas, with a temperature
of $\sim300$ K. The high density gas ($10^4$ cm$^{-3} \lesssim n \lesssim 10^5$
cm$^{-3}$) does not cool below $\sim 148$ K at any time.
The re-expanding outer layers of the cloud are significantly colder with
$T\approx60$ K, because of adiabatic expansion.
The turbulence that can been seen in the passing shock is due to the thin-shell instability.
During the shock-cloud interaction, the clump mass has increased from 19 M$_\odot$ to 40 M$_\odot$. We do not expect this clump to be gravitationally unstable as the minimum Jeans mass is 1000 M$_\odot$ (Figure \ref{fig_clump}).

After the passage of the shock the dense cloud is embedded in the high pressure, hot, low density interior of the supernova remnant. Our simulations do not have the spatial resolution to resolve the boundary layer between these two phases (we also do not include thermal conduction or model the external irradiation of the cloud), so the details of the boundary layer are probably not very reliable. The dominant physical process, however, is the simple pressure confinement of the cloud, and this is well-captured by our calculation.  By the time the cloud is accelerated off the simulation domain it is entering an equilibrium phase of a pressure-confined cloud, similar to the cometary phase for irradiated clouds \citep{BerMcK90}.

\section{Discussion} \label{sec:dicusssion}

\begin{center}
\begin{table*}

 {\begin{tabular}{|c||c|c|c|c|c|c|c|c|}
 Model No. & Supernova Energy & Ambient cloud  & H\,\textsc{ii} region & Clump density & Temperature & Clump distance & Shock velocity & Clump fate \\
 &($10^{51}$ erg)& density (cm$^{-3}$) &included &(cm$^{-3}$)&of clump (K)&(pc) &(km s$^{-1}$)& \\
  \hline \hline 
M01 &10 & 10& Yes&100& 872 &46& 39& compressed clump\\
M02 & 2.0 & 10& Yes&100& 872  &46&-&shell stalled\\
M03 &1.0 & 10& Yes&100& 872  &46&- &shell stalled\\
M04 & 0.6 &10& Yes&100& 872  &46& - &shell stalled\\
M05 &10 & 1& No&100 & 200 &50&200&destroyed\\
M06 & 2.0 & 1& No&100 &200 &50&46& small fragments\\
M07 &1.0 &  1& No&100 &200 &50&26& destroyed\\
M08 & 0.6 & 1 & No&100 &200 &50&16& destroyed\\
M09 & 1.0 & 1 & No &100 &200 &40&49& small fragments\\
\hline
  \end{tabular}}

\caption{ This table presents the initial conditions of a number of shock-cloud models and the corresponding fate of the clump at the end of the simulation. There are four end states of the clump: i)  the clump is unaffected by the shock as the shell stalled before reaching the clump, ii) the clump is fully compressed into a single core, iii) the clump fragments into smaller dense pieces and iv) the clump no longer exists and is destroyed.} \label{model-table}
\end{table*}
\end{center}

We have made a first investigation of
the importance of
non-equilibrium cooling processes occurring at all temperatures in
primordial cloud-shock interactions (i.e.\ a SN shell interacting with
primordial gas at redshift $z = 10$).
This is an interesting case to study in its own right, for predicting
the minimum mass that a metal-free star could potentially have.
It is also the limiting case of considering shock-cloud interactions at
extremely low metallicity, and so is useful for establishing a control
simulation, against which models with non-zero metallicity can later be
compared (Dhanoa et al., in prep).
The progenitor gas cloud for the hyper metal-poor star SDSS
J102915+172927 \citep{Caffau11} (with a total metallicity $Z\lesssim 10^{-5}$ Z$_\odot$)
may have formed in a similar environment that was metal-free, but which
became slightly polluted with supernova ejecta.

We include non-equilibrium chemistry to capture the radiative cooling occurring during the interaction of a shock and a small cloud, to establish if it is possible to form low-mass stars via this method. Considering a primordial chemistry for this process may be a simplification; because metals from the supernova ejecta would interact and mix within the shell once the discontinuity between the shell and the ejecta is disrupted by the impact of the clump \citep{Tenorio96}. However, the metallicity of the shell is expected to be near zero \citep{Salvaterra04} and according to \citet{CenR} the shock velocity ensures that the clump remains mostly unaffected by metals. If this is true then modelling the shock and cloud as metal-free is a good approximation.

 We calculate the minimum Jeans mass of the of the compressed clump with only H$_2$ cooling (i.e.\ the minimum possible), and therefore represent a limiting case for shock-cloud interactions
for both low-metallicity models and primordial models which include deuterium cooling.
We find that the fractional abundance of H$_2$ in the high density region exceeds $3\times10^{-3}$, hence deuterium cooling may become important in this interaction \citep{Nakamura02}. On the other hand, \citet{WolHai11} have shown that HD cooling is supressed by UV radiation fields that are five orders of magnitude weaker than what is required to supress H$_2$ cooling.
Thus we expect that models with no HD cooling are applicable to a wider range of environments than models with HD cooling, once stars have begun forming in the vicinity.


We assume that the progenitor star is formed in a dark matter halo that is large enough so that edge effects do not need to be taken into account for a radius of $r \leq 50$ pc. \citet{Vasiliev08} highlighted an important link between the radial distribution of primordial gas prior to the supernova explosion and the subsequent evolution of the supernova remnant; the state of the supernova shell directly influences the formation of extremely metal-poor stars.
This distribution is heavily dependent on the size of the
H\,\textsc{ii} region prior to the star's explosion.
Studies which reproduce the abundance patterns in extremely metal-poor stars by modelling the evolution and explosion of metal-free stars stars
\citep{Nomoto06, Joggerst09, Joggerst10}, suggest that metal-poor stars are formed
by metal-free stars within a mass range of $15- 40$ M$_\odot$. The explosion mechanism for metal-free stars is uncertain, especially above 30 M$_\odot$, and so the star can have a range of explosion energies from $0.6 - 10 \times 10^{51}$ erg, which are associated with core collapse supernovae and hypernovae.

A clump initially at distance $r \ge 40$ pc  from the star can safely be assume to be neutral, because  Figure \ref{fig_10cc_HII} shows that the clump does not interact with any ionising radiation. Clouds found closer to the progenitor star may evaporate, or at a minimum, have a different thermal state to a neutral cloud. Radiation between $11.18-13.6$ eV photodissociates H$_2$ molecules and so has a knock-on heating effect on the gas. This dissociation radiation propagates further than ionising radiation, and without any dust present we expect that clump is completely atomic in the pre-supernova stage. In the 2D model dissociative photons from the hot gas is assumed to be negligible \citep{Vasiliev08}, but the possible effects of UV radiation on the clump should be investigated in more detail in future work.

After exploring a number of explosion energies (see models M01$-$M04 in Table \ref{model-table}), we found that only the shock formed from a hypernova explosion ($10^{52}$ erg) reached and compressed the clump. When extending our study by exploring other ambient cloud densities (models M05$-$M09 in Table \ref{model-table}), it emerges that the shock velocity determines the fate of the neutral clump. If the shock is too fast the clump is destroyed. When the supernova shock is too slow, the clump is only slightly compressed but inevitably destroyed. This is because the initial shock causes a secondary shock to travel through the rest of the clump, finally the gas disperses and flows downstream with the supernova shock. We therefore find that a small range of shock velocities ($30-50$ km s$^{-1}$) which can cause the clump to compress or fragment. Here the cooling time is equal to or less than the collapse/compression time and the velocity of the shock causes at least half of the clump to be compressed. Shock velocities above 40 km s$^{-1}$ cause the clump to fragment into smaller clumps, while below this velocity we find the clump is compressed.

The clump is near a supernova remnant so it will be exposed to cosmic
rays, but the cosmic ray spectrum and intensity is unknown because of
uncertainties in the expected interstellar magnetic field and the
explosion mechanism for metal-free stars. We have assumed that the spectrum with be close to the observed spectrum in the Galactic environment, in keeping with \citet{Stacy07}. 
In this model we include a background cosmic ray ionization rate of $10^{-18}$ s$^{-1}$, as this rate was found to produce an overall cooling effect.
We have not explored X-rays in this work, which would be produced by the supernova remnant.
This would increase the H$_2$ abundance of gas ahead of the shell by increasing the free electron content \citep{Ferrara98, Haiman97} and should be subject to further investigation.
The effects of a range of cosmic ray ionization rates ($10^{-18}-10^{-15}$ s$^{-1}$) and their associated heating on the shock-clump interactions will also be explored future work.

The shocked clump of model M01 implodes because of the passage of the supernova shock (Figure\,\ref{fig_10cc_SN}). This is the same behaviour seen in 3D simulations of clouds interacting with clumps \citep{Melioli06, Leao09, Johansson13}, and earlier 2D work \citep[e.g.][]{KleMcKCol94}. We find that in our simulation the clump gains a maximum density of $\sim 78000$ cm$^{-3}$, which is a density enhancement of $10^{2.89}$ but does not become Jeans unstable. \citet{Vaidya13} show that self-gravity has no effect on the clump at this point of the shock interaction, where the implosion is pressure-driven and the clump reaches its maximum density. This gives us confidence that the implosion phase is correctly captured by our simulation. \citet{Johansson13} investigate the compression of a $n=17$\,cm$^{-3}$ cloud (with radius 1.5 pc)  and find higher densities enhancements of $10^3 - 10^5$. They also conclude that the clump will not become Jeans unstable. It is worth noting that their work considers solar metallicity gas with an equilibrium cooling function. Hence this may change when the model is refined to include non-equilibrium cooling. 

Dust is assumed to be the major coolant in low-metallicity environments \citep{Klessen12, Schneider12}. How quickly it can form in a primordial supernova ejecta and the extent of mixing that would occur during this cloud-shock interaction are still open questions. It is believed that dust is quickly destroyed in the reverse shocks formed when the supernova shell begins to travel within the multiphase neutral medium \citep{Cherchneff10, Silvia10}. Without much dust in the environment, we cannot expect metal-line cooling to drastically lower the Jean mass, especially
at metallicities $\le10^{-5}\,\rm{Z}_\odot$. In light of this, much further work is required to investigate the effects of cosmic rays and external radiation fields (especially X-ray and UV) on this process, because there may be important positive feedback effects \citep{Ricotti02, Oshea05} that have not been considered so far. 

\section{Conclusion} \label{sec:conc}
We have presented a metal-free shock-cloud model, which simulates a supernova remnant interacting with a cloud at redshift $z=10$.   We model a dense clump ($n=100$ cm$^{-3},\, r =1.3$ pc) embedded in a 10 cm$^{-3}$ ambient cloud, which is 40 pc from the progenitor star. We consider realistic pre-supernova conditions by including the effects of stellar radiation from a 40 M$_\odot$ metal-free star on the multi-phase neutral medium. At the end of the star's main-sequence lifetime, a hypernova ($10^{52}$ erg) is initiated and the evolution of the supernova shell and its subsequent interaction with the dense clump is studied. Radiative cooling is a crucial process in the shock-cloud interaction, allowing the formation of dense cold gas that may be susceptible to gravitational collapse. During this process we have comprehensively modelled the radiative (non-equilibirum) cooling taking place. 

We followed the evolution of the supernova remnant and its interaction with the surrounding ionised and neutral medium. When the radiative shell interacts with the metal-free clump, it reaches a maximum of density $\sim78000$ cm$^{-3}$. This is a $10^{2.89}$ density enhancement and is consistent with Galactic shock-cloud models considering solar metallicity gas with equilibrium cooling functions. The clump undergoes  a reduction in Jeans mass from $10^5$ M$_\odot$ to $10^3$ M$_\odot$, but does not become gravitationally unstable. Further work is required to ascertain the effect of cosmic rays, X-rays and UV radiation on the clump during the supernova phase.

In this work, we found an optimal range of shock velocities ($30- 50$ km s$^{-1}$) which compress small metal-free clouds. Below this range the cloud is slightly perturbed by the supernova shock and is not subject to any appreciable density enhancement. Above this range the clumps are destroyed, therefore the results  by \citet{MacBroHer03} are overoptimistic, as they assume the cloud survives a 200 km s$^{-1}$ interaction. 

In this initial study we have only considered a single clump with fixed
radius and density, varying the supernova energy and the density of the
medium that the clump is embedded in.
We have shown that the Jeans mass is indeed reduced significantly by the
shock-cloud interaction, but not sufficiently to form stars with $<1$ M$_\odot$.
In order to draw more general conclusions about the possibility of
forming such low mass stars from metal free gas, we plan to follow up
this work by considering a range of clump sizes and central densities.

When investigating model M01, we have achieved an appreciable Jean mass reduction of a small dense clump and a density enhancement comparable to Galactic studies, by including non-equilibrium metal-free radiative cooling. Further refinement of this model by including low-metallicity chemistry
plus positive feedback effects from cosmic rays, X-rays and UV radiation, may cause a further reduction in Jeans mass. Galactic models should be extended to include non-equilibrium cooling, as this work has shown that it is the dominant process in shock-cloud interactions. 

\section*{Acknowledgements}
The authors would like to thank S.C.O. Glover and J.M.C. Rawlings for their helpful discussions. This work used the DiRAC Complexity system, operated by the University of Leicester IT Services, which forms part of the STFC DiRAC HPC Facility (www.dirac.ac.uk). This equipment is funded by BIS National E-Infrastructure capital grant ST/K000373/1 and  STFC DiRAC Operations grant ST/K0003259/1. DiRAC is part of the National E-Infrastructure. JM acknowledges funding during this project by a fellowship from the Alexander von Humboldt Foundation and from the Deutsche Forschungsgemeinschaft priority program 1573, ``Physics of the Interstellar Medium''.
\bibliography{mf_letter}
 \appendix
\section{Chemistry Network}\label{appen:chem}
The full chemical network is displayed in Table \ref{chem_network}.
All the molecular reaction rates (R07 -R42) have been adapted for the temperature range ($10-10^{9}$ K) have been divided into two categories: i) Formation rates  (listed in Table \ref{Formation_molcules}) and ii) Destruction rates (listed in Table  \ref{Extended_molcules}). \bigskip

Most of the UMIST 06 rates are valid until 41,000K. If a formation rate is valid up to a lower temperature, the value at the maximum temperature range is kept constant for temperatures above until 41,000K. Above 41,000 K all formation rates are cut-offand the reaction rates take on the the following forms:
\begin{eqnarray*}
K_1&=&k \times \rm{exp}\left(1.0 -\frac{T}{41000.0}\right)\\ 
K_2&=&k \times \rm{exp}\left(10 \times\left(1.0 -\frac{T}{41000.0}\right)\right)
\end{eqnarray*}
where $k$ is the value of the rate at 41000K. The details of how each formation reaction is treated, can be found in Table \ref{Formation_molcules}.\bigskip

The destruction rates are extrapolated above their valid temperature range. Above this temperature, if there is a maximum value after which the rate decreases (T$_{ex}$), this maximum value is kept constant for all higher temperatures (T $>$ T$_{ex}$). All the destruction rates, with the corresponding maximum extrapolation temperatures and temperatures ranges are displayed in Table \ref{Extended_molcules}. 

\onecolumn
\begin{longtable}{l l r}
  \hline \hline
Reaction No. &Reaction & References for rate coefficients\\
 \hline
 \endhead
 R01 & H\p + e\m $\rightarrow$ H + $\gamma$             & H\\
 R02 & He\p + e\m $\rightarrow$ He + $\gamma$         & VF\\
 R03 & He\p\p + e\m $\rightarrow$ He\p + $\gamma$   & VF\\
 R04 & H + e\m $\rightarrow$ H\p + e\m + e\m              & V\\
 R05 & He + e\m $\rightarrow$ He\p + e\m + e\m          & V\\
 R06 & He\p + e\m $\rightarrow$ He\p\p + e\m + e\m    & V\\
 R07 & H$_2$ + H $\rightarrow$ H + H + H                   & GA08\\
 R08 & H\m + H $\rightarrow$ H + H + e\m                    & GA08\\
 R09 & H\m + He $\rightarrow$ He + H + e\m                & GA08\\
 R10 & H$_2$ + H$_2$ $\rightarrow$ H$_2$ + H + H   & UM06\\
 R11 & H$^-$ + e$^-$ $\rightarrow$ H + e$^-$ + e$^-$ & JR\\
 R12 & H$_2$ + He$^{+}$ $\rightarrow$ He + H$^+$ + H & UMO6\\
 R13 & H$_2$ + e$^-$ $\rightarrow$ H + e$^-$ + H      & UM06\\
 R14 & \htp + e$^-$ $\rightarrow$ H$^+$ + e$^-$ + H   & R14*\\
 R15 & HeH\p + e\m $\rightarrow$ He$^+$ + e$^-$ + H & R14*\\
 R16 & H$^+$ + H $\rightarrow$ \htp + $\gamma$        & UM06, GA08\\
 R17 & \hp + He $\rightarrow$ HeH\p + $\gamma$       & UM06\\
 R18 & H + e$^-$ $\rightarrow$ H$^-$ + $\gamma$      & UM06, GA08\\
 R19 & HeH\p + e\m $\rightarrow$ He + H                    & UM06\\
 R20 & \htp + e$^-$ $\rightarrow$ H + H                        & UM06\\
 R21 & $\hthp$ + e$^-$ $\rightarrow$ H + H + H           & UM06\\
 R22 & $\hthp$ + e$^-$ $\rightarrow$ H$_2$ + H          & UM06\\
 R23 & H\m + \htp $\rightarrow$ H + H + H                   & GA08\\
 R24 & H + He$^{+}$ $\rightarrow$ He + H$^+$           & UM06,hd\\
 R25 & H$_2$ + He$^{+}$ $\rightarrow$ He + \htp       & UM06\\
 R26 & \hp + H$^{-}$ $\rightarrow$ H + H                     & UM06\\
 R27 & H$^-$ + \htp $\rightarrow$ H$_2$ + H              & UM06\\
 R28 & H$^-$ + He$^+$ $\rightarrow$ He + H              & UM06\\
 R29 & H + \htp $\rightarrow$ H$_2$ + \hp                   & UM06\\
 R30 & \htp + H$_2$ $\rightarrow$ $\hthp$ + H           & UM06\\
 R31 & H$^-$ + $\hthp$ $\rightarrow$ H$_2$ + H$_2$  & UM06\\
 R32 & H + HeH$^+$ $\rightarrow$ He + \htp              & UM06\\
 R33 & H$_2$ + HeH\p $\rightarrow$ He + $\hthp$     & UM06\\
 R34 & \htp + He $\rightarrow$ HeH\p + H                   & UM06\\
 R35 & H\m + H\p $\rightarrow$ \htp + e\m                  & SK87\\
 R36 & H$^-$ + H $\rightarrow$ H$_2$ + e$^-$           & UM06\\
 R37 & H + CR $\rightarrow$ \hp + e$^-$                     & UM06\\
 R38 & He + CR $\rightarrow$ He\p + e$^-$                &UM06\\
 R39 & H$_2$ + CR $\rightarrow$ \hp + H + e$^-$      &UM06\\
 R40 & H$_2$ + CR $\rightarrow$ H + H                     &UM06\\
 R41 & H$_2$ + CR $\rightarrow$ \hp + H$^-$            &UM06\\
 R42 & H$_2$ + CR $\rightarrow$ \htp + e$^-$            &UM06 \\
 \hline\hline
 \\
 \caption{Metal free chemistry network: \\
 References- UM06 =UMIST database for astrochemistry [rate 06, non-dipole enhanced] \citep{Woodall07}; GA08 = \citet{Glover08}; H =\citet{Hummer94}; GP98 = \citet{Galli98}; SK87= \citet{Shapiro87}; hd = matching scheme; R14*= same value as R14; JR= private communication with Jonathan Rawlings; V=\citet{Voronov97}; VF= \citet{Verner96}} \label{reactiontable}\label{chem_network}
 \end{longtable}

\begin{table}
\begin{center}
 \begin{tabular}{|c||c|c|c|c|}
 \hline
Reaction & Valid Temperature    & Below Minimum & Above Maximum  & Cut off Type\\
Number & Range (K)  &Temperature &  Temperature & T $> $41000K \\
 \hline\hline
 R16 & S:10 -- 32000& -  & C & CT2\\
 R17 & 16 -- 100    & C & E & CT\\
 R18 & S:10 -- 41000& - & - & CT2\\
 R29 & 10 -- 41000& - & - & CT\\
 R30 & 10 -- 41000     & - & - & CT\\
 R33 & 10 -- 41000     & - & - & CT\\
 R34 & 10 -- 41000     & - & - & CT\\
 R35 & 10 -- 41000     & - & - & CT\\
 R36 & S:10 -- 41000  & - & - & CT\\
\hline
  \end{tabular}
\end{center}
\caption{Molecular reactions that are cut-off at 41000 K:  E= rate
extrapolated; C= max/min value kept constant and extended; - = Not
Applicable; S= switching between different reaction rates within temperature range;
CT2= $k\exp{(10.0\times(1.0-\rm{T}/41000))}$ and CT=
$k\exp{(1.0-\rm{T}/41000})$ are exponential cut-off for T> 41000K and $k$ is the value
of the reaction rate at 41000 K } \label{Formation_molcules}
\end{table}

\begin{table}
\begin{center}
\begin{tabular}  {|c||c|c|c|c|}
  \hline
Reaction & Valid Temperature    & Below  & Above   & Maximum Extrapolation\\
Number & Range of Rate (K)  & Range&   Range&  Temperature T$_{ex}$ (K)\\
 \hline\hline
 R07 & 1833 -- 41000 & E & E & $10^9$\\
 R08 & 10 -- 10000 & - & C & -\\
 R09 & 10 -- 10000   & - & C & -\\
 R10 & 2803 -- 41000 & E & E & $10^7$ \\
 R11 &  10 -- 41000& - & E & $10^5$ \\
 R12 & 100 -- 300 & E & E & $10^8$\\
 R13 & 3400 --41000 & E & E & $10^8$\\
 R14 & 3400 --41000 & E & E & $10^8$ \\
 R15 & 3400 --41000 & E & E & $10^8$\\
 R19 & 10 -- 300     & - & E & $10^9$\\
 R20 & 10 -- 300     & - & E & $10^9$ \\
 R21 & 10 -- 1000    & - & E & $10^9$\\
 R22 & 10 -- 1000    & - & E & $10^9$\\
 R23 & 10 -- 10000   & - & C & -\\
 R24 & S:10 -- 41000 & - & C & -\\
 R25 & 10 -- 300     & - & E & $10^9$\\
 R26 & 10 -- 300     & - & E & $10^4$\\
 R27 & 10 -- 300     & - & E & $10^9$\\
 R28 & 10 -- 300     & - & E & $10^9$\\
 R31 & 10 -- 300     & - & E & $10^9$\\
 R32 & 10 -- 41000   & - & E & $10^9$\\
 R37& 10 -- 41000   &- & C & -\\
 R38& 10 -- 41000   &- & C & -\\
 R39& 10 -- 41000   &- & C & -\\
 R40& 10 -- 41000   &- & C & -\\
 R41& 10 -- 41000   &- & C & -\\
 R42& 10 -- 41000   &- & C & -\\
 \hline
 \end{tabular}
 \end{center}
 \caption{ Molecular reactions adapted to maximum temperature ($10^{9}$ K): E= rate is extrapolated to a maximum Extrapolation Temperature
(T$_{ex}$) and then extended as a constant after that temperature;
C= max/min value kept constant; - = Not Applicable; S= a number of reaction rates utilised
within temperature range} \label{Extended_molcules}
 \end{table}
\twocolumn
\section{Cooling test}\label{app:cooltest}

\begin{figure}
\centering{
\includegraphics[scale=0.35, angle =270]{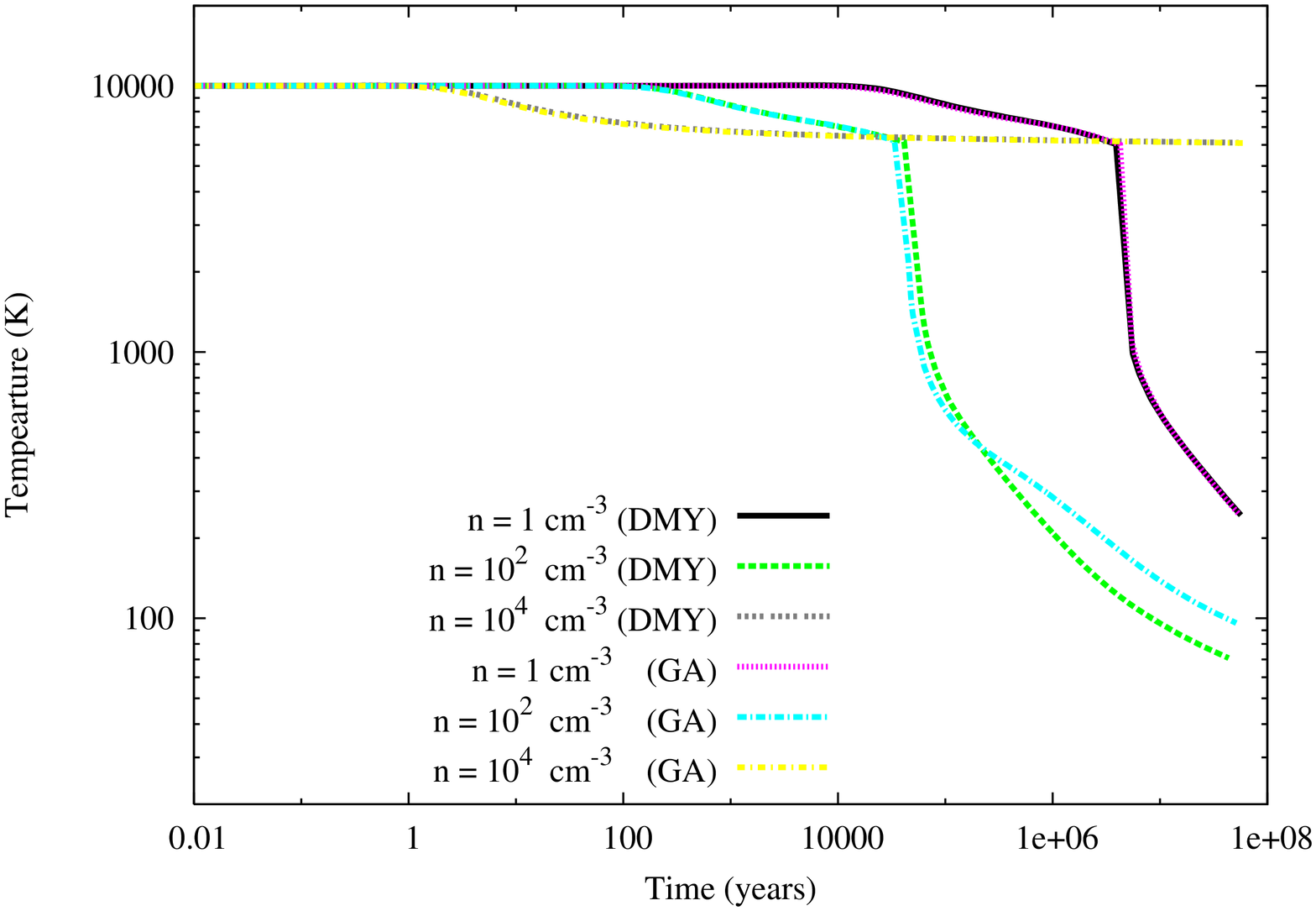}
}
\caption{One-zone test to compare the microphysics module (DMY) against the chemistry network presented by \citet{Glover08} (GA). {\changed Both networks use the same cooling rates for each species.} The gas is initially fully ionised and has a temperature of $10^4$ K. Three different densities are investigated: $n=1$ cm$^{-3}$, $n=100$ cm$^{-3}$ and $n=10^4$ cm$^{-3}$}\label{chem_test}
\end{figure}

Fig.~\ref{chem_test} displays a comparison of the primordial chemistry network presented in this work (DMY) and that of \citet[][GA]{Glover08} which includes 32 reactions that contain hydrogen and helium species only. {\changed The GA deuterium reactions are not included}. Notably GA have included three-body reactions and density dependent reactions for:
\begin{eqnarray*}
&\rm{H}_2 + H_2 \rightarrow H + H +H_2\\
&\rm{H}_2 + He \rightarrow H + H +He.
\end{eqnarray*}  
These reactions have been neglected in our network. However, \citet{Glover08} do not include H$_3^+$ and HeH$^+$.

In this test we adopt a one-zone constant density model, where both chemistry networks are linked to the {\changed same set of cooling functions, i.e. the H$_2$ and H$_2^+$ cooling functions  provided by \citet{Glover08} and \citet{Hollenbach79} plus the atomic cooling functions given by \citet{Fukugita94}, \citet{Hummer94}, \citet{Shapiro87} and \citet{Peebles71}}. The initial temperature of the gas is $10^4$ K and three densities are investigated: $n=1\,\cm$, $n=100\,\cm$ and $n=10^4\,\cm$. The gas is allowed to chemically evolve and cool over $5\times10^7$ years.
 
For the low density test (i.e. $n = 1\, \cm$) both microphysics modules reach the same
temperature of 244 K. In the test for $n = 10^4\, \cm$ the temperatures are very close; our module
cools down to 6115 K and the GA module cools to 6090 K. At this density the H$_2$ cooling is
within the LTE regime. The temperatures are very close, the difference is due to the rates we
have included and not because the three body reactions were excluded. Three body reactions are
dominant for densities $n \ge10^5\, \cm$, and can be neglected as we do not expect the densities in the
module to reach this value. Finally for $n = 100\, \cm$ we obtain 69 K whilst the GA module
obtains 90 K. When we include the reactions that are missing from our module, we still obtain 69
K. This highlights that the differences in temperature are due to the differences between the rates
used in UMIST 06 database and the GA module.

\section{1D supernova shell expansion}\label{app:SNtest}
\begin{figure}
 \includegraphics[width=0.49\textwidth]{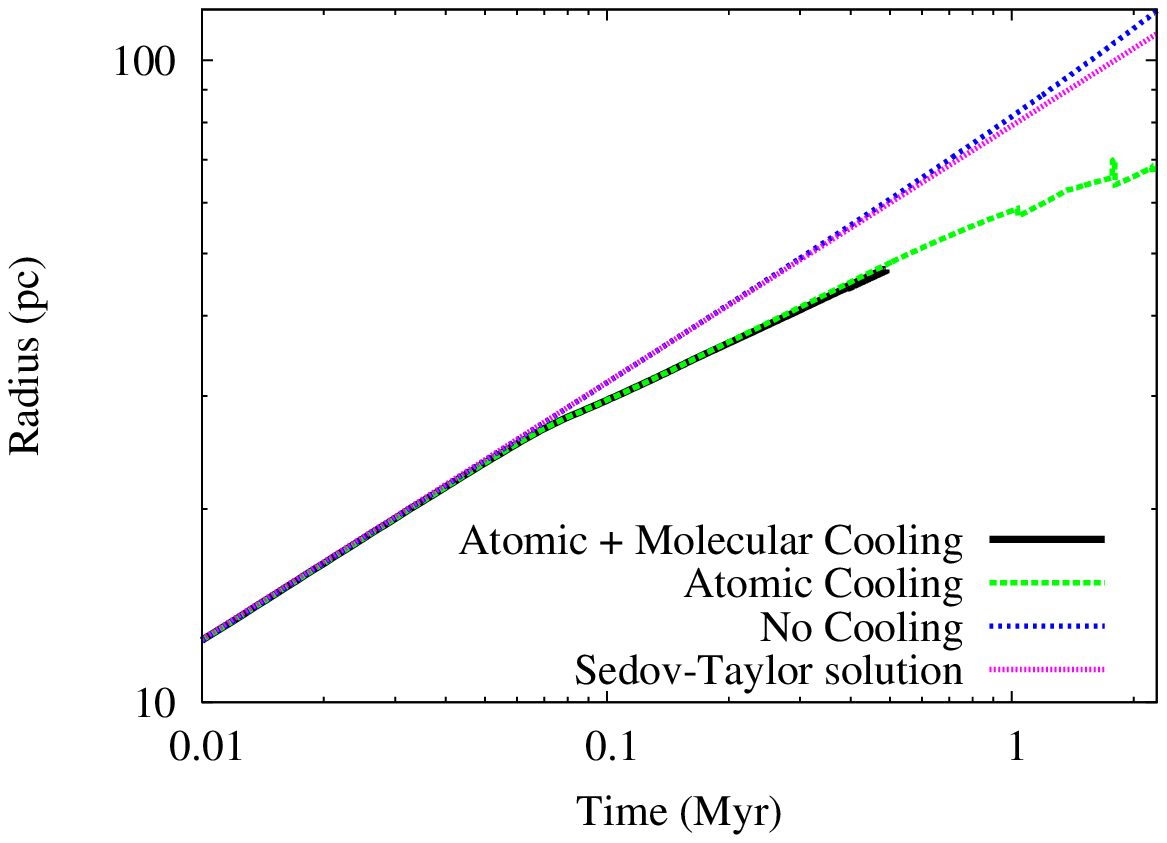}
 \includegraphics[width=0.49\textwidth]{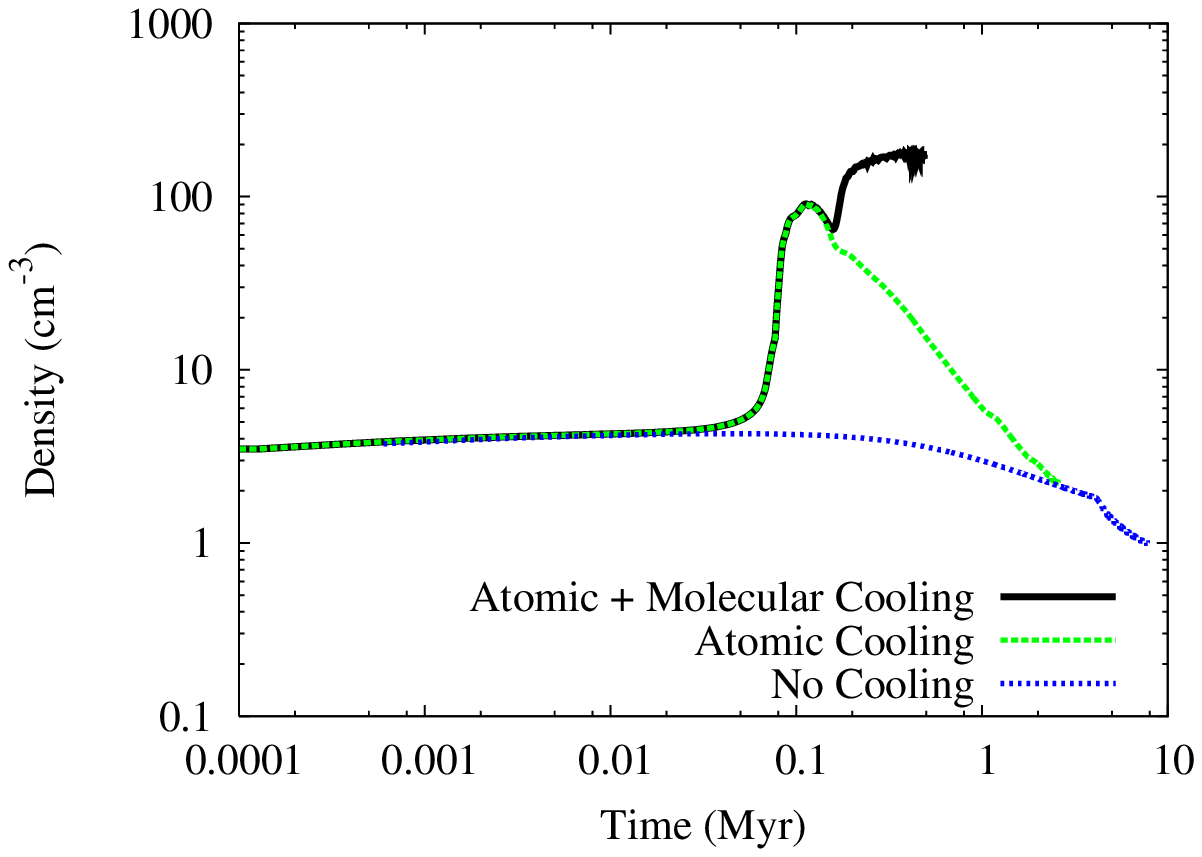}
  \caption{
    Supernova shell expansion as a function of time for an adiabatic calculation, a calculation with atomic line cooling only, and a calculation with atomic and molecular cooling switched on.
    The expansion radius is compared to the analytic Sedov-Taylor solution in the upper plot.
    The lower plot shows the maximum gas number density in the shell as a function of time for the same three models.
  }
  \label{fig_SNexp}
\end{figure}
Fig.~\ref{fig_SNexp} shows the results of a 1D test, in which the expansion of a blastwave is followed using different chemistry/cooling assumptions: adiabatic with no chemistry, including chemistry but only atomic coolants, and including chemisty with atomic and molecular coolants.
The radius of the SN forward shock (upper panel) and maximum density in the shell (lower panel) are plotted as a function of time since explosion.
We used uniform radial grid with 5120 grid zones between $r=0$ and $r=130$\,pc, and input $10^{51}$\,ergs of thermal energy in the 8 grid zones closest to the origin.
The ISM is a constant density medium with $\rho=2.44\times10^{-24}\,$g\,cm$^{-3}$ at a redshift of 20.
The initial ISM temperature is $T=10^4$\,K (corresponding to a pressure of $p\approx1.5\times10^{-12}$\,dyne\,cm$^{-2}$).
Without any cooling this can be compared to the Sedov-Taylor solution, and when cooling and chemistry are included we compare to the results of \citet{Machida05}.

The adiabatic calculation matches the Sedov-Taylor solution until about 0.8\,Myr, after which the shock runs ahead of this solution.
The explanation for this is that the shock weakens as it slows down at late times, and the ISM ambient pressure is no longer negligible.
This breaks the scale-free nature of the analytic solution, and the result is that the shock radius advances faster than predicted at late times \citep[cf.][]{RagCanRodEA12}.

At about 0.05\,Myr the simulations with cooling start to decelerate and deviate from the adiabatic solution.
The expansion rate changes from the Sedov-Taylor value $R_\textrm{sh}\propto t^{2/5}$ to the momentum-conserving value $R_\textrm{sh}\propto t^{1/4}$.
Atomic cooling is initially much stronger than molecular cooling, so both of these runs match each other until the molecular cooling begins to affect the shell and the ISM at $t\approx0.2$\,Myr.
At later times the shell density in the cooling model decreases steadily because it can no longer cool, and the weak forward shock keeps adding lower entropy gas to the shell.
The molecular cooling model has a higher density shell once molecular cooling becomes important at $t\approx0.2$\,Myr, because it can cool to much lower temperatures.
This has the further effect that the shell remains at a high density for much longer.

The molecular cooling calculation shows that we get compression factors of $>100\times$ in the shell at $t\geq0.2\,$Myr.
This model disagrees strongly with \citet{Machida05} (see their fig.~4), who found only weak density increase in the supernova shell for times up to $10^7$ years.
The density in their analytic model was set by the imposed pressure-confining boundary conditions on the shell, so we suspect that one of the boundary conditions was incorrect.

\label{lastpage}

\end{document}